\documentclass[aps,pra,twocolumn,superscriptaddress,amsmath,amssymb]{revtex4}
\usepackage{graphicx}
\usepackage{amsmath}
\usepackage{color}
\usepackage[normalem]{ulem}
\def\cp#1{\mathbf{#1}}

\begin{document}

\title{Pairing Superfluidity in Spin-Orbit Coupled Ultracold Fermi Gases}
\author{Wei Yi}
\email{wyiz@ustc.edu.cn}
\affiliation{Key Laboratory of Quantum Information, University of Science and Technology of China,
CAS, Hefei, Anhui, 230026, People's Republic of China}
\affiliation{Synergetic Innovation Center of Quantum Information and Quantum Physics, University of Science and Technology of China, Hefei, Anhui 230026, China}
\author{Wei Zhang}
\email{wzhangl@ruc.edu.cn}
\affiliation{Department of Physics, Renmin University of China, Beijing 100872, People's Republic of China}
\affiliation{Beijing Key Laboratory of Opto-electronic Functional Materials and Micro-nano Devices,
Beijing 100872, People's Republic of China}
\author{Xiaoling Cui}
\email{xlcui@iphy.ac.cn} \affiliation{Beijing National Laboratory
for Condensed Matter Physics, Institute of Physics, Chinese Academy
of Sciences, Beijing, 100190, People's Republic of China}

\date{\today}
\begin{abstract}
We review some recent progresses on the study of ultracold Fermi gases with synthetic spin-orbit coupling. In particular, we focus on the pairing superfluidity in these systems at zero temperature. Recent studies have shown that different forms of spin-orbit coupling in various spatial dimensions can lead to a wealth of novel pairing superfluidity. A common theme of these variations is the emergence of new pairing mechanisms which are direct results of spin-orbit-coupling-modified single-particle dispersion spectra. As different configurations can give rise to single-particle dispersion spectra with drastic differences in symmetry, spin dependence and low-energy density of states, spin-orbit coupling is potentially a powerful tool of quantum control, which, when combined with other available control schemes in ultracold atomic gases, will enable us to engineer novel states of matter.
\end{abstract}
\maketitle

\section{Introduction}\label{introduction}

The recent experimental realization of synthetic gauge field in ultracold atomics gases has greatly extended the horizon of quantum simulation in these systems~\cite{gauge1exp,gaugenist1,gaugenist2,gauge2exp,fermisocexp1,fermisocexp2,shuaiexp,engelsexp,nistfesh,zjnatphys,shuainatphys,chenyong}. A particularly important case is the implementation of synthetic spin-orbit coupling (SOC), a non-Abelian gauge field, in these systems, where the internal degrees of freedom of the atoms are coupled to the atomic center-of-mass motional degrees of freedom~\cite{gauge2exp,fermisocexp1,fermisocexp2,nistfesh,zjnatphys,chenyong}. In condensed-matter materials, SOC plays a key role in many interesting phenomena, such as the quantum spin Hall effects, topological insulators, and topological superconductors~\cite{kanereview,zhangscreview,alicea}. Although the form of the synthetic SOC currently realized in cold atoms differs crucially from those in condensed-matter systems, there exist various theoretical proposals on realizing synthetic SOC which can induce topologically nontrivial phases~\cite{Rashba_Spielman_1,Rashba_Spielman_2,Rashba_Xu_1,Rashba_Liu,Rashba_Spielman_3,Rashba_Xu_2,congjun3d,spielman_3d_1}. Thus, the hope of simulating the various topological phases, the topological superfluid state in particular, in the highly controllable environment of an ultracold atomic gas stimulated intensive theoretical studies on spin-orbit coupled Fermi gases~\cite{zhang,sato,gongzhang,2d2,wy2d,2d1,wmliu,helianyi,tfflo0,tfflo1,tfflo2,tfflohu,chuanwei3dsoc,huhuintFF,huhui3dsoc}. Furthermore, recent studies suggest that other exotic superfluid phases and novel phenomena can be engineered with carefully designed configurations~\cite{iskin,thermo,melo,polaronwy,xiaosen,iskinnistsoc,puhan3dsoc,xiangfa3dsoc,wyfflo,wyfflolong,cuiyipairing,trimer1,trimer2}. As such, SOC has a great potential of becoming a powerful tool of quantum control in ultracold atomic gases.

In this review, we focus on the zero-temperature pairing physics in a spin-orbit coupled ultracold Fermi gas. We will discuss the exotic superfluid phases in systems with different spatial dimensions and with different forms of SOC. A fundamentally important effect of SOC is the modification of the single-particle dispersion spectra~\cite{hzsocbec,cplwu,soc3,soc4}. We will start from there and show how this effect leads to interesting pairing phases such as the topological superfluid state, the various gapless superfluid states, the SOC-induced Fulde-Ferrell (FF) state, and the topological FF state. We will also touch upon the topic of exotic few-body states in spin-orbit coupled Fermi systems whose stability also benefits from the SOC-modified single-particle dispersion.

The paper is organized as follows: in Sec.~\ref{sec_implementation}, we briefly introduce the implementation scheme of SOC in current cold atom experiments. In Sec.~\ref{sec_singlespec}, we present the single-particle dispersion spectra under various forms of SOC. Then in Sec.~\ref{sec_pairing}, we analyze the general pairing mechanism in these systems based on the modified single-particle dispersion, and present the exotic superfluid phases and the rich phase diagrams under different configurations. We further discuss the possibilities of engineering novel pairing states and novel few-body states in Sec.~\ref{sec_engineer}. Finally, we summarize in Sec.~\ref{sec_fin}.

\section{Implementing synthetic gauge field}\label{sec_implementation}

The principle behind most of the proposals for an artificial gauge potential is based on the adiabatic theorem and the associated geometrical phase~\cite{dalibardreview}. In general, by engineering the atom-laser interaction, the atoms experience an adiabatic potential when moving through space. The resulting geometrical phase appearing in the effective Hamiltonian gives rise to the artificial gauge potential. To see this, we start from the full Hamiltonian
\begin{equation}
H=H_0+V[\mathbf{r}(t)],
\end{equation}
where $H_0=\mathbf{P}^2/2m$ is the kinetic energy and $V[\mathbf{r}(t)]$ describes the atom-laser coupling, whose spatial dependence is related to the atomic motion.

Formally, let us expand the wave function at any given time $|\Psi(\mathbf{r},t)\rangle$ onto the eigen basis $\{|\phi_{\alpha}(\mathbf{r})\rangle\}$ of $V(\mathbf{r})$
\begin{equation}
|\Psi(\mathbf{r},t)\rangle=\sum_{\mathbf{\alpha}}c_{\alpha}(\mathbf{r},t)|\phi_{\alpha}(\mathbf{r})\rangle,
\end{equation}
where $c_{\alpha}$'s are the time-dependent expansion coefficients. Substituting the expansion above into the time-dependent Schr\"{o}dinger's equation and projecting it into the subspace of the ${\alpha}$-th eigen state, we have
\begin{equation}
i\hbar\frac{\partial}{\partial t}c_{\alpha}(\mathbf{r},t)=E_{\alpha}(\mathbf{r})c_{\alpha}+\sum_{\beta}\left\langle\phi_{\alpha} |H_0c_{\beta}(\mathbf{r},t)|\phi_{\beta}\right\rangle,\label{eqnadiabaticfull}
\end{equation}
where $E_{\alpha}(\mathbf{r})$ satisfies $V(\mathbf{r})|\phi_{\alpha}(\mathbf{r})\rangle=E_{\alpha}(\mathbf{r})|\phi_{\alpha}(\mathbf{r})\rangle$.
Assuming the adiabatic condition, under which the slow center-of-mass motion of an atom adiabatically follows the fast internal dynamics governed by $V(\mathbf{r})$, we may retain only $\beta=\alpha$ in Eq. (\ref{eqnadiabaticfull}) to get
\begin{equation}
i\hbar\frac{\partial}{\partial t}c_{\alpha}(\mathbf{r},t)=E_{\alpha}(\mathbf{r})c_{\alpha}+\left\langle\phi_{\alpha} |H_0c_{\alpha}|\phi_{\alpha}\right\rangle,
\end{equation}
which effectively describes the motion of an atom in the adiabatic potential $E_{\alpha}(\mathbf{r})$. To make the geometrical phase stand out, we further examine the term involving the kinetic energy
\begin{eqnarray}
\langle \phi_{\alpha}|H_0c_{\alpha}|\phi_{\alpha}\rangle&=&\frac{1}{2m}\langle \phi_{\alpha}|\mathbf{P}\left\{\sum_{\beta}|\phi_{\beta}\rangle\langle\phi_{\beta}|\left[\mathbf{P}c_{\alpha}|\phi_{\alpha}\rangle\right]\right\} \nonumber\\
&=&\frac{1}{2m}\left(\mathbf{P}-\mathbf{A}\right)^2c_{\alpha}+Wc_{\alpha}.
\end{eqnarray}
Here, the geometrical vector potential $\mathbf{A}=i\hbar\left\langle\phi_{\alpha}|\nabla\phi_{\alpha}\right\rangle$, and the geometrical scalar potential $W=\sum_{\beta\neq\alpha}\hbar^2\left|\langle\phi_{\beta}|\nabla\phi_{\alpha}\rangle \right|^2/2m$. Hence, apart from an energy shift due to the scalar potential $W$, the effective Hamiltonian for an atom in the adiabatic potential $E_{\alpha}(\mathbf{r})$ can be written as
\begin{equation}
H_{\text{eff}}=\frac{1}{2m}\left(\mathbf{P}-\mathbf{A}\right)^2+E_{\alpha}(\mathbf{r}).
\end{equation}

\begin{figure}[tbp]
\centering
\includegraphics[width=9cm]{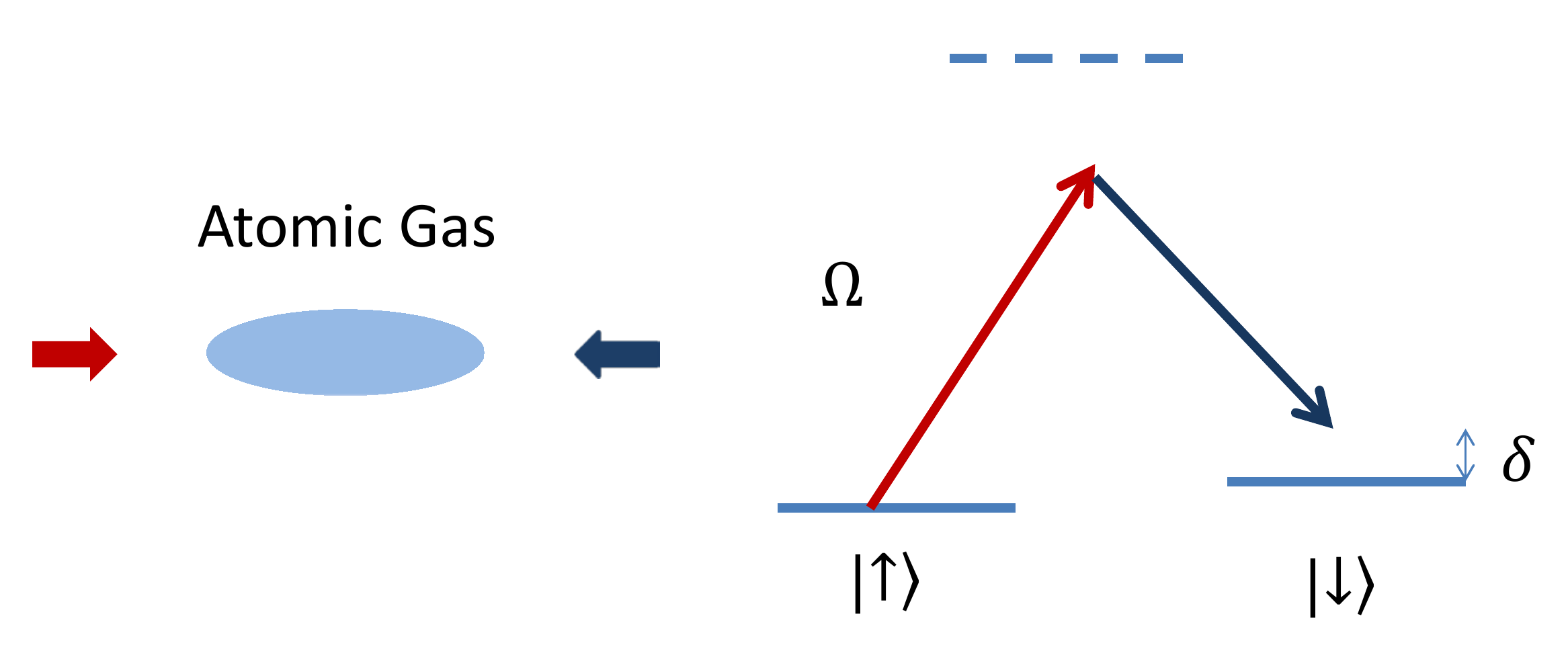}
\caption{Schematics on the experimental implementation of SOC in cold atoms. $\Omega$ is the effective Rabi frequency of the Raman process, $\delta$ is the two-photon detuning. The pseudo-spin states ($|\uparrow\rangle,|\downarrow\rangle$) are hyperfine states within the ground state hyperfine manifold.}\label{fig_levelscheme}
\end{figure}

The physical implication is just what we have stated at the beginning, for a particle moving in an adiabatic potential, its external motion adiabatically follows the internal dynamics at each spatial location. As a result, the internal states of the particle may change as the particle is moving through space. When the change in internal states involves only a phase factor, the gauge potential associated with the geometrical phase is Abelian, as is the case for synthetic electromagnetic fields. When the change in internal states involves a general rotation in the Hilbert space spanned by the internal states, the gauge potential associated with the geometrical phase can be non-Abelian, as is the case for synthetic SOC.

Experimentally, the adiabatic potential is typically generated by coupling the internal states of an atom with lasers, and it was Spielman's group at NIST that had first realized a uniform vector gauge potential in a BEC of $^{87}$Rb atoms~\cite{gauge1exp}. The scheme utilizes Raman lasers to couple hyperfine states in the ground state manifold of $^{87}$Rb, such that when an atom jumps from one internal state to another via the Raman process, its center-of-mass momentum also changes. Applying a similar scheme, vortices were later observed in a BEC of $^{87}$Rb with synthetic magnetic field, followed by the implementation of synthetic electric field~\cite{gaugenist1,gaugenist2}. In 2010, via a slightly modified Raman scheme (see Fig.~\ref{fig_levelscheme}), Spielman's group was able to generate synthetic SOC, a non-Abelian version of the synthetic gauge field, in a BEC of $^{87}$Rb atoms~\cite{gauge2exp,xjlgauge,nistsoctheory}. This was soon followed by the realization of synthetic SOC in ultracold Fermi gases by the ShanXi group and the MIT group in 2012~\cite{fermisocexp1,fermisocexp2}. In the following, we will address the currently realized synthetic SOC in cold atomic gases as the NIST SOC.

In the case of the NIST SOC, the effective single-particle Hamiltonian can be written as
\begin{eqnarray}
H&=&\sum_{\cp k}\frac{\hbar^2}{2m}({\cp k}+k_0{\cp x}\sigma_z)^2-\frac{\delta}{2}\sigma_z+\frac{\Omega}{2}\sigma_x\\
&\Rightarrow& \frac{\hbar^2}{2m}({\cp k}+k_0{\cp x}\sigma_x)^2-\frac{\delta}{2}\sigma_x-\frac{\Omega}{2}\sigma_z \label{eqn_nistsoc}
\end{eqnarray}
where we have rotated the spin basis in the second line so that the effective Hamiltonian takes the form of an equal mixture of Rashba $(k_x\sigma_x+k_y\sigma_y)$ and Dresselhaus $(k_x\sigma_x - k_y\sigma_y)$ SOC. At the moment, only the effective one-dimensional SOC in the form of Eq. (\ref{eqn_nistsoc}) has been implemented experimentally. Note that in Eq. (\ref{eqn_nistsoc}), the effective Rabi-frequency $\Omega$ and the two-photon detuing $\delta$ can be seen as Zeeman fields perpendicular and parallel to the direction of SOC, respectively.

\section{Single-particle dispersion under SOC}\label{sec_singlespec}

A fundamental effect of SOC in ultracold Fermi gases is the modification of the single-particle dispersion. Under SOC, both the symmetry and the low-energy density-of-states are different, leading to unconventional BEC in Bose systems, and exotic pairing states in Fermi systems with attractive interactions. In this section, we will focus on the single-particle dispersion under typical forms of SOC, and discuss their potential impact on many-body systems.

\subsection{Rashba SOC}

The single-particle Hamiltonian under the Rashba SOC can be written as
\begin{equation}
H=\sum_{\mathbf{k},\sigma}\epsilon_{\mathbf{k}}a^{\dag}_{\mathbf{k},\sigma}a_{\mathbf{k},\sigma}+\sum_{\mathbf{k}} \left[\alpha\left(k_x-ik_y\right)a^{\dag}_{\mathbf{k},\uparrow}a_{\mathbf{k},\downarrow}+H.C.\right],
\end{equation}
where $\epsilon_{\cp k}=\hbar^2k^2/2m$, $a^{\dag}_{\mathbf{k},\sigma}$ ($a_{\mathbf{k},\sigma}$) is the creation (annihilation) operator for the pseudo-spin $\sigma=\{\uparrow,\downarrow\}$, $\alpha$ is the SOC strength, and $H.C.$ stands for Hermitian conjugate. The pseudo-spin here is related to the adiabatic potential, and its exact relation with the hyperfine states is scheme dependent. The Hamiltonian can be diagonalized by introducing the annihilation operators in the so-called helicity basis
\begin{eqnarray}
a_{\mathbf{k},+}&=&\frac{1}{\sqrt{2}}\left(e^{i\varphi_{\mathbf{k}}}a_{\mathbf{k},\uparrow}+a_{\mathbf{k},\downarrow}\right),
\\ a_{\mathbf{k},-}&=&\frac{1}{\sqrt{2}}\left(e^{i\varphi_{\mathbf{k}}}a_{\mathbf{k},\uparrow}-a_{\mathbf{k},\downarrow}\right),
\end{eqnarray}
with $\varphi_{\cp k}=\arg (k_x+ik_y)$. The diagonalized Hamiltonian
\begin{equation}
H=\sum_{\mathbf{k},\lambda=\pm}\xi_{\lambda}a^{\dag}_{\mathbf{k},\lambda},
\end{equation}
where $\xi_{\pm} = \epsilon_{\bf k} \pm \alpha k$ are the resulting single-particle eigen energies.
\begin{figure}[tbp]
\centering
\includegraphics[width=9cm]{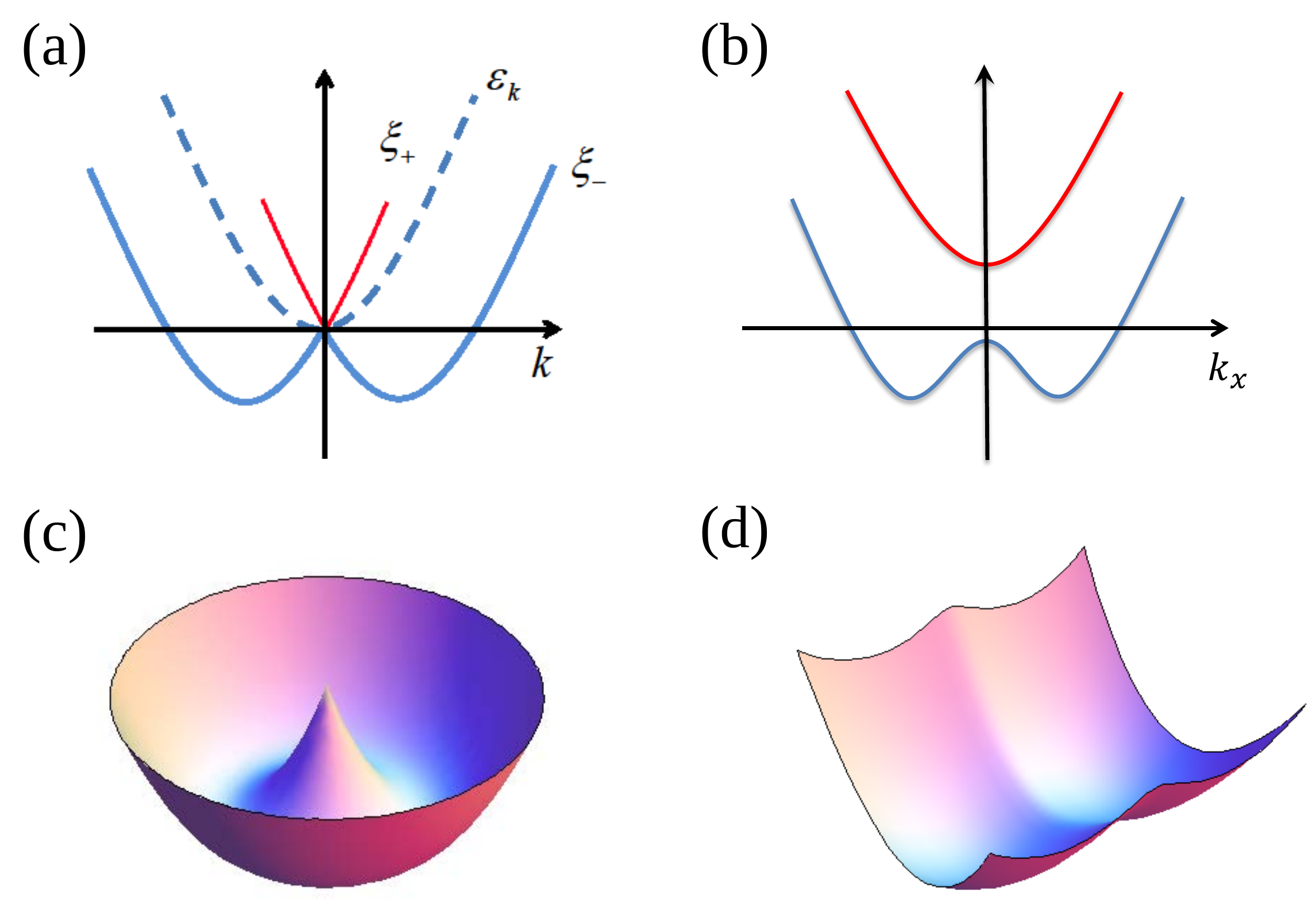}
\caption{Illustration of the single-particle dispersion spectra under SOC. (a) The dashed curve represents the free-particle spectra ($\epsilon_{\cp k}$); the solid curves are the spectra for the helicity branches ($\xi_{\pm}$). (b) The spectra of the helicity branches under SOC and an out-of-plane Zeeman field $(h \sigma_z)$. (c) The single-particle spectrum of the lower helicity branch under the Rashba SOC, three-dimensional view. (d) The single-particle spectrum of the lower helicity branch under the NIST SOC, three-dimensional view.}\label{fig_specsoc}
\end{figure}

The single-particle spectra under Rashba SOC is illustrated in Fig.~\ref{fig_specsoc}, where it is clear that SOC breaks the inversion symmetry and splits the spin spectra into two helicity branches. Due to the symmetry of the Rashba SOC, points of the lowest energy in the lower helicity branch form a ring in momentum space, and the ground state is infinitely degenerate in this case [see Fig.~\ref{fig_specsoc}(c)]. Correspondingly, the density-of-states for a Rashba spin-orbit coupled spectra in three dimensions is a constant at the lowest energy. This peculiar single-particle spectra and density of states naturally lead to interesting many-body phenomena~\cite{hzsocbec,cplwu,soc3,soc4}.

For noninteracting bosons, as the ground-state degeneracy is infinite, BEC is no longer possible at zero temperature even in three dimensions. However, a weak interaction can induce a spontaneous symmetry breaking, and the bosons will condense to either one point or two opposite points on the degenerate ring in momentum space, depending on the interaction parameters. This leads to the so-called plane-wave phase and the stripe phase in a uniform interacting Bose gas with Rashba SOC~\cite{hzsocbec,cplwu}.

For a noninteracting Fermi gas at zero temperature, atoms occupy all the low-energy states up to the Fermi energy. As the atom number density increases, the topology of the Fermi surface changes and the system undergoes a Lifshitz transition. Furthermore, it has been shown that for atoms with attractive $s$-wave interaction between the spin species and with Rashba SOC, the two-body bound state energy is enhanced due to the increased density of states at low energies~\cite{soc3,soc4}. In fact, a two-body bound state exists in three dimensions even in the weak-coupling limit, similar to the case of a two-dimensional problem without SOC, suggesting an effective reduction of dimensions, which is consistent with the SOC-modified density of states.

\subsection{NIST SOC}

\begin{figure}[tbp]
\centering
\includegraphics[width=9cm]{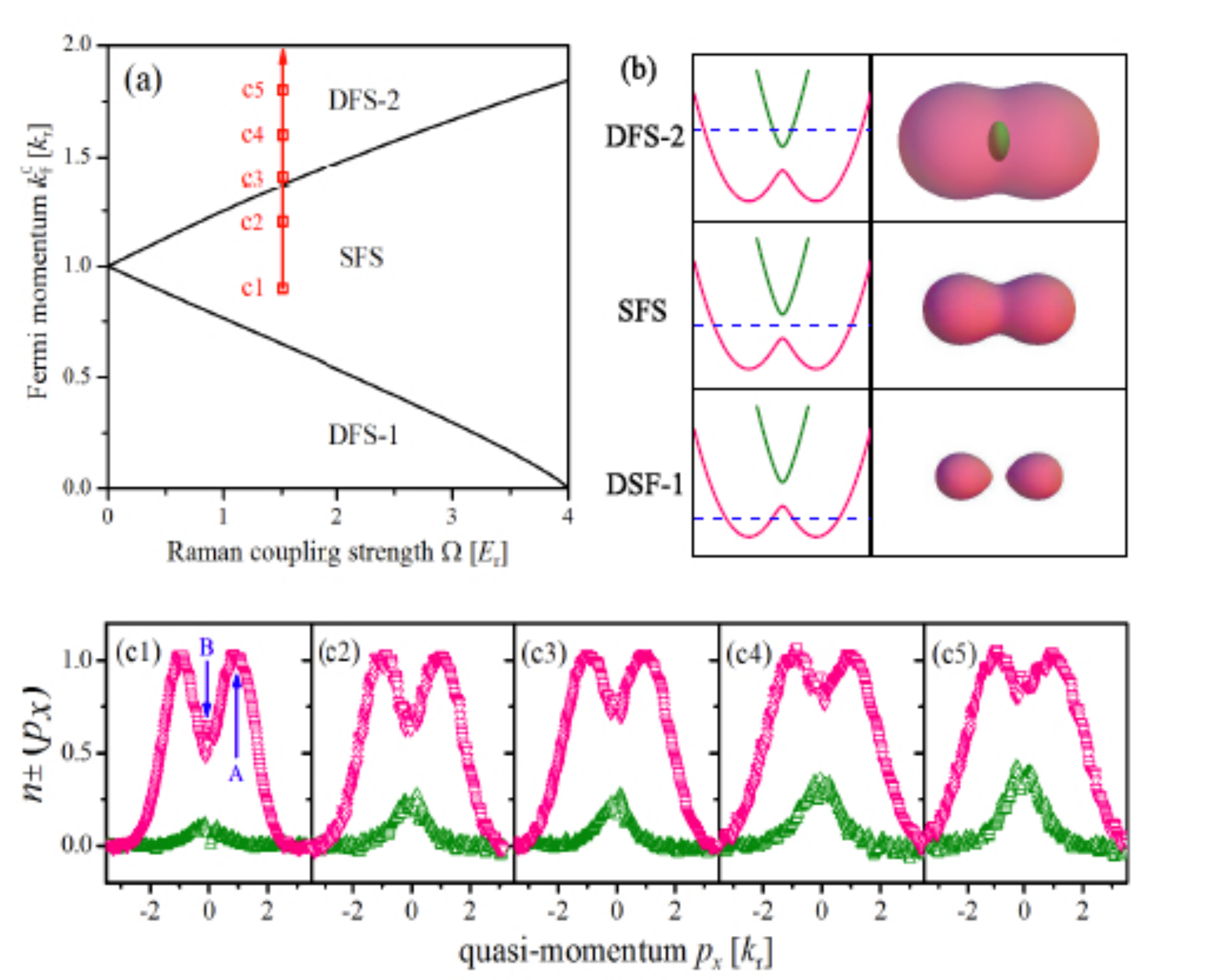}
\caption{Experimental observation of the Lifshitz transition for a noninteracting degenerate Fermi gas with spin-orbit coupling \cite{fermisocexp1}. Reprinted with permission from the American Physical Society.}\label{zjsocexp}
\end{figure}

The single-particle Hamiltonian under the NIST SOC can be written as
\begin{align}
H=&\sum_{\mathbf{k},\sigma}\epsilon_{\mathbf{k}}a^{\dag}_{\mathbf{k},\sigma}a_{\mathbf{k},\sigma}+\sum_{\mathbf{k}} h\left( a^{\dag}_{\mathbf{k},\uparrow}a_{\mathbf{k},\downarrow}+H.C.\right)\nonumber\\
&+\sum_{\mathbf{k}}\alpha k_x\left(a^{\dag}_{\mathbf{k}\uparrow}a_{\mathbf{k}\uparrow}-a^{\dag}_{\mathbf{k}\downarrow}a_{\mathbf{k}\downarrow}\right),
\end{align}
where $a^{\dag}_{\mathbf{k},\sigma}$ ($a_{\mathbf{k},\sigma}$) is the creation (annihilation) operator for different hyperfine states. Compared with Eq. (\ref{eqn_nistsoc}), the SOC parameter $\alpha$ is related to the momentum transfer of the Raman process, and the effective Zeeman field $h$ is proportional to the effective Rabi-frequency. The helicity basis that diagonalizes the Hamiltonian becomes
\begin{align}
a_{\cp k,\pm}=\pm\beta_{\cp k}^{\pm}a_{\cp k,\uparrow}+\beta_{\cp k}^{\mp}a_{\cp k,\downarrow},
\end{align}
with $\beta^{\pm}_{\cp k}=\big[\sqrt{h^2+\alpha^2k_x^2}\pm \alpha k_x \big]^{1/2}/\sqrt{2} [ h^2+\alpha^2k_x^2]^{1/4}$. The single-particle dispersion of the helicity branches $\xi_{\pm}=\epsilon_{\cp k}\pm\sqrt{h^2+\alpha^2k_x^2}$. When $h<m\alpha^2/\hbar^2$, the degenerate ground state manifold consists of two points in momentum space, in contrast to a ring under the Rashba SOC [see Fig.~\ref{fig_specsoc}(d)]. The spin components of the helicity branches are momentum dependent under the NIST SOC, which is also different from that of the Rashba SOC. Although these differences can lead to drastically different many-body effects, as we will show later, cold atomic gases under the NIST SOC still preserve many of the interesting properties of those under the Rashba SOC. For example, in a BEC under the NIST SOC, the stripe phase and the plane-wave phase can still be identified in the finite-temperature phase diagram~\cite{shuainatphys}. While in a noninteracting degenerate Fermi gas, topological changes of the Fermi surface have been observed experimentally (see Fig.~\ref{zjsocexp}).

\subsection{Other forms of SOC}

Besides the Rashba and the NIST SOC, the isotropic SOC of the form $\mathbf{k}\cdot\mathbf{\sigma}$, or the three-dimensional SOC, has also been discussed recently by various authors~\cite{congjun3d,spielman_3d_1,chuanwei3dsoc,huhui3dsoc,puhan3dsoc,xiangfa3dsoc,congjun3dsoc,soc3}. For bosons, spin textures of the ground state under an isotropic SOC may acquire interesting topology. For fermions, it is expected that the system under appropriate parameters can behave like three-dimensional topological insulators or Weyl semimetals. Additionally, the application of an effective Zeeman field can deform the single-particle dispersion spectra along the direction of the field. As we will see later, this would lead to novel pairing states under attractive inter-particle interactions. Other more exotic forms of synthetic SOC exist and have been discussed in the literature~\cite{ngoldman1,ngoldman2,huhui3comp}. A common feature of these SOCs is the absence of counterparts in naturally occurring condensed-matter systems, which makes the quantum simulation of systems with exotic forms of SOC in ultracold atomic gases more appealing.

\section{Pairing physics under SOC}\label{sec_pairing}

With the understanding of single-particle dispersions under SOC, we now examine the pairing physics in an attractively interacting Fermi gas with SOC.
We start with the conventional BCS pairing mechanism in the weak-coupling limit, which should help us to appreciate the fundamentally new pairing mechanisms in a spin-orbit coupled Fermi gas.

\subsection{Interbranch pairing and intrabranch pairing}

\begin{figure}[tbp]
\centering
\includegraphics[width=9cm]{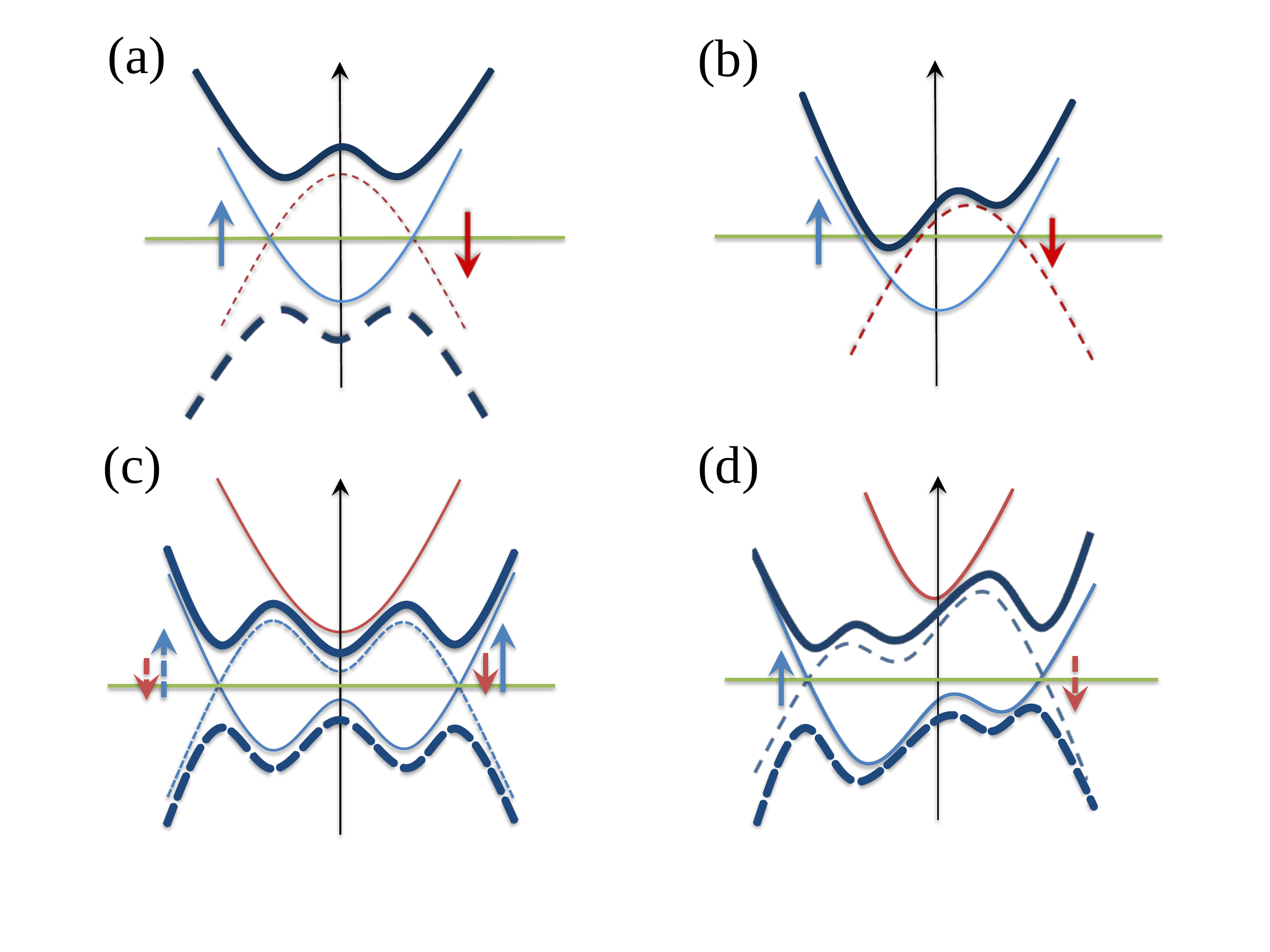}
\caption{Schematics of the pairing mechanisms. (a) Illustration of the BCS pairing mechanism. (b) Illustration of the conventional FF pairing mechanism. (c) Pairing mechanism under SOC and an out-of-plane Zeeman field. (d) Pairing mechanism of the SOC-induced FF state with both out-of-plane and in-plane Zeeman fields. In all figures, the thin red (blue) curve represents the particle (hole) dispersion of the noninteracting system, the thick solid (dashed) curve represents the quasiparticle (quasihole) dispersion under the pairing interaction. The horizontal green line is the Fermi energy. See Ref.~\cite{wyfflo}}\label{fig_bcspairing}
\end{figure}

It is well known that a two-component Fermi sea without spin imbalance becomes unstable in the presence of a small attractive interaction. The resulting Cooper instability leads to a BCS ground state, which is the basis for understanding conventional superconductivity in most metals at low enough temperatures. In ultracold Fermi gases, pairing superfluidity has been experimentally observed thanks to the Feshbach resonance technique~\cite{bcsbecreview,chinchengreview}. The effective BCS mean-field Hamiltonian of such a system can be written as
\begin{align}
H-\mu N&=\sum_{\cp k}
\begin{bmatrix}a^{\dag}_{\cp k\uparrow}&a_{-\cp k\downarrow}\end{bmatrix}
\begin{bmatrix}
\epsilon_{\cp k}-\mu & \Delta\\
\Delta^{\ast} & -(\epsilon_{\cp k}-\mu)
\end{bmatrix}
\begin{bmatrix}a_{k\uparrow}&a^{\dag}_{-{\cp k}\downarrow}\end{bmatrix}\nonumber\\
&+\sum_{\cp k}(\epsilon_{\cp k}-\mu)-\frac{|\Delta|^2}{U},\label{eqn_bcspairing}
\end{align}
where $U$ is the bare interaction rate, $\Delta=U\langle a_{-\cp k\downarrow}a_{\cp k\uparrow}\rangle$ is the pairing order parameter, and $\mu$ is the chemical potential, which, in the weak-coupling limit, is essentially the Fermi energy of the system. Under the Hamiltonian (\ref{eqn_bcspairing}), the BCS pairing mechanism in the weak-coupling limit can be schematically illustrated as Fig.~\ref{fig_bcspairing}(a). The dispersion of the spin-up particles crosses that of the spin-down holes at the Fermi surface, where the pairing mean field $\Delta$ couples the two branches, leading to an avoided crossing and the quaiparticle (quasihole) dispersions. From the positions of the avoided crossings, we see that the pairing state in this case has zero center-of-mass momentum. In the presence of a spin imbalance, or an effective Zeeman field, the Fermi surface mismatch between the two spin species can lead to competition between the BCS pairing mechanism and the Fulde-Ferrell-Larkin-Ovchinnikov pairing mechanism~\cite{fflo}. In the latter case, the dispersion of the minority spin species is shifted so that pairing can take place near the Fermi surface [see Fig.~\ref{fig_bcspairing}(b)], thus leading to a finite center-of-mass momentum pairing state. In both of these cases, an important feature is that pairing involves both the particle and the hole branches of different spin species. We address this kind of pairing mechanism as interbranch pairing. Note in Fig.~\ref{fig_bcspairing}(b) and in the rest of the review, we consider the simple case where pairing occurs with a single-valued center-of-mass momentum, the so-called Fulde-Ferrell (FF) state.

Under SOC, with different single-particle dispersions, a new pairing mechanism emerges, which becomes more transparent and also physically more interesting in the presence of effective Zeeman fields. Extending the BCS mean-field theory, we may write down the effective mean-field Hamiltonian of a two-component Fermi gas under the Rashba SOC and an effective out-of-plane Zeeman field
\begin{align}
H_{\rm{eff}}=&\frac{1}{2}\sum_{\cp k}
\begin{bmatrix}
\xi_{\cp k}+h&0&\alpha k e^{i\varphi_{\cp k}}&\Delta\\
0&-(\xi_{\cp k}+h)& -\Delta^{\ast} & \alpha k e^{-i\varphi_{\cp k}}\\
\alpha k e^{-i\varphi_{\cp k}}&-\Delta&\xi_{\cp k}-h&0\\
\Delta^{\ast}&\alpha k e^{i\varphi_{\cp k}}&0&-(\xi_{\cp k}-h)
\end{bmatrix}\nonumber\\
&+\sum_{\cp k}(\epsilon_{\cp k}-\mu)-\frac{|\Delta|^2}{U},
\end{align}
where $h$ is the effective Zeeman field along the $z$ direction. The Hamiltonian above is written the pseudo-spin basis $\left\{a_{\mathbf{k}\uparrow},a^{\dag}_{-\mathbf{k}\uparrow},a_{\mathbf{k}\downarrow},a^{\dag}_{-\mathbf{k}\downarrow}\right\}^{T}$. Under SOC, the mean-field ground state of the system has both $s$- and $p$-wave pairing components, as both $\sum_{\cp k}\langle a_{-\cp k\downarrow}a_{\cp k\uparrow}\rangle$ and $\sum_{\cp k}\langle a_{-\cp k\sigma}a_{\cp k\sigma}\rangle$ ($\sigma=\uparrow,\downarrow$) are finite~\cite{2d2,soc4}. Starting from the Hamiltonian above, we now examine the pairing mechanism in the weak-coupling limit.

In the absence of any Zeeman fields, as illustrated in Fig.~\ref{fig_specsoc}(a), the two helicity branches cross at the origin in momentum space. Under an effective out-of-plane Zeeman field, i.e., when $h$ is finite, the helicity branches are coupled and a gap opens up at the origin [see Fig.~\ref{fig_bcspairing}(c)]. When the chemical potential lies in this gap, as is the case in Fig.~\ref{fig_bcspairing}(c), pairing can occur within the lowest helicity branch, leading to the opening of pairing gaps at the crossings of the particle and hole dispersions of the lower branch. This is possible since SOC mixes different spin species so that $s$-wave pairing between spin-up and spin-down atoms can happen within the same helicity branch. We call this pairing mechanism intrabranch pairing. As we will demonstrate in the next section, for a Rashba SOC, this is the pairing scenario that results in the exotic topological superfluid state. When the chemical potential lies above the gap, or when the interaction becomes stronger, interbranch pairing becomes important and competes with intrabranch pairing. From the location of the avoided crossings, we see that pairing states in this case have zero center-of-mass momentum.

When an in-plane Zeeman field is added, the single-particle dispersion spectra become deformed. As shown in Fig.~\ref{fig_bcspairing}(d), in this case, intrabranch pairing naturally leads to pairing states with finite center-of-mass momenta. Apparently, these SOC-induced FF pairing states are the result of interplay between SOC and Fermi surface asymmetry, and are different in mechanism from the conventional FF (cFF) states discussed previously.

With these qualitative analysis in the weak-coupling limit, we have already seen the various possibilities of exotic pairing states under SOC. In the following, let us see some more concrete examples.

\subsection{Topological superfluid state and gapless superfluid state}

Perhaps the most important exotic pairing state in a spin-orbit coupled Fermi system is the topological superfluid state in two dimensions, originally investigated in the context of semiconductor/superconductor heterstructures with Rashba SOC, $s$-wave pairing order and an out-of-plane Zeeman field~\cite{kanereview,zhangscreview,alicea}. As has been shown in Ref.~\cite{zhang,sato}, SOC breaks the inversion symmetry and mixes up different spin components in the helicity branches. The introduction of the external Zeeman field breaks the time-reversal symmetry and opens up a gap between different helicity branches. In the weak-coupling limit, when the Fermi surface of the system lies in this gap, the ground state of the system becomes a topological superfluid state with Majorana zero modes on the boundaries or at the cores of vortex excitations. Importantly, these non-Abelian Majorana zero modes are protected by a bulk gap and are therefore useful for fault-tolerant topological quantum computation~\cite{nayak}.

The stabilization of the topological superfluid state can be understood as the direct result of intrabranch pairing, which leads to a mixture of $s$- and $p$-wave symmetries in the pairing order parameter~\cite{2d2,soc4}. In parameter regions where the $p$-wave pairing dominates, the $p_x\pm ip_y$ symmetry of the pairing order parameter inherited from the Rashba SOC allows one to map the system onto a $p$-wave superfluid with the corresponding symmetry~\cite{sato}.

\begin{figure}[tb]
\begin{center}
\includegraphics[width=8cm]{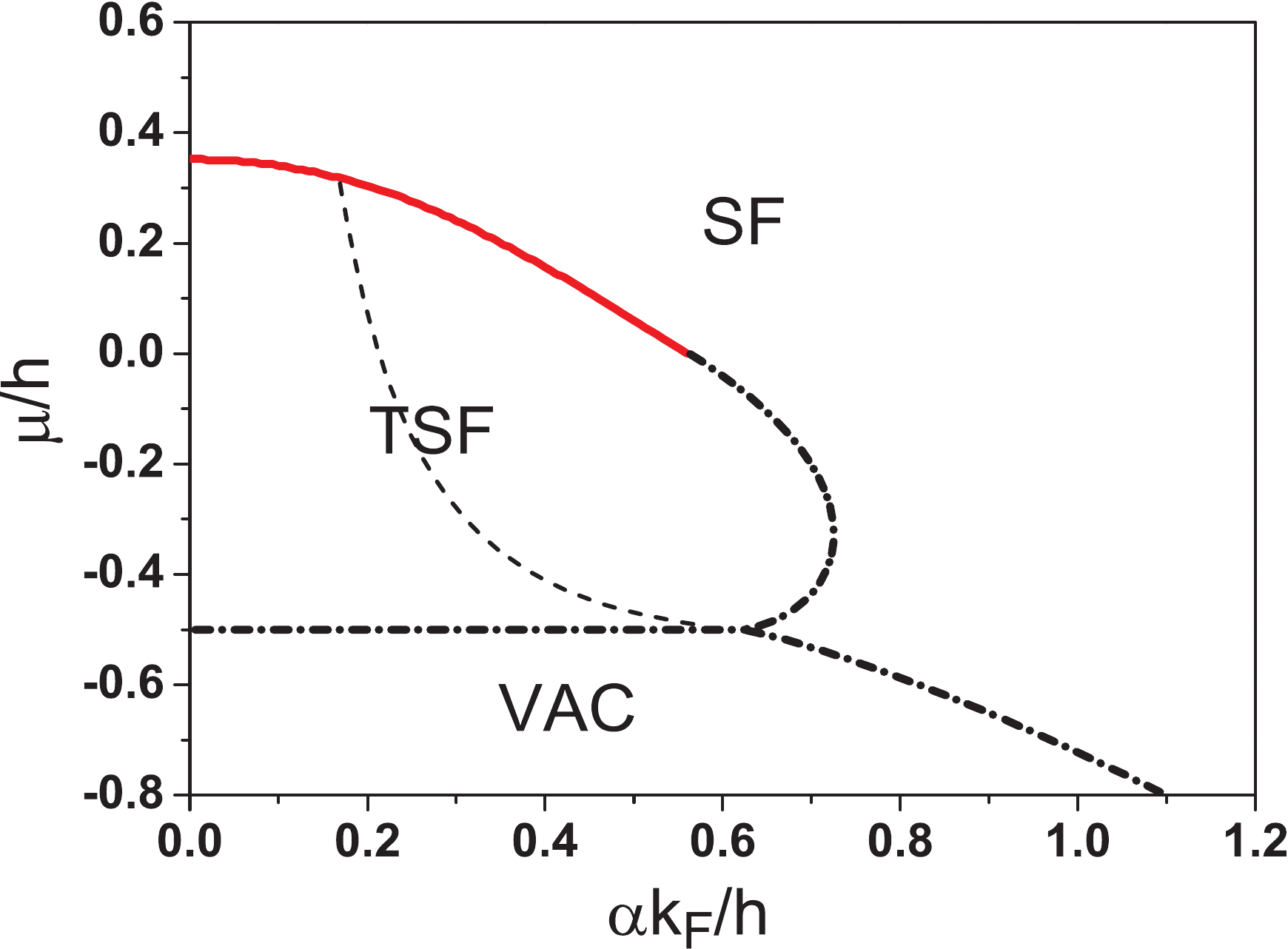}
\end{center}
\caption{Typical phase diagram in the $\alpha$-$\mu$ plane for a two-dimensional Fermi gas under Rashba SOC and an out-of-plane Zeeman field $h$.
The first-order phase transition is shown in red solid curve while the second-order phase transitions are shown in dash-dotted black curves.
The thin dashed curve in the topological superfluid (TSF) region marks the $\Delta/h=10^{-3}$ threshold. The effective Zeeman field $h$ is taken to be the energy unit, while the unit of momentum $k_h$ is defined through $\hbar^2k_h^2/(2m)=h$. Here, VAC represents the vacuum. See Ref.~\cite{wy2d}.}
\label{fig_Eb}
\end{figure}

\begin{figure*}[tbp]
\centering
\includegraphics[width=5cm]{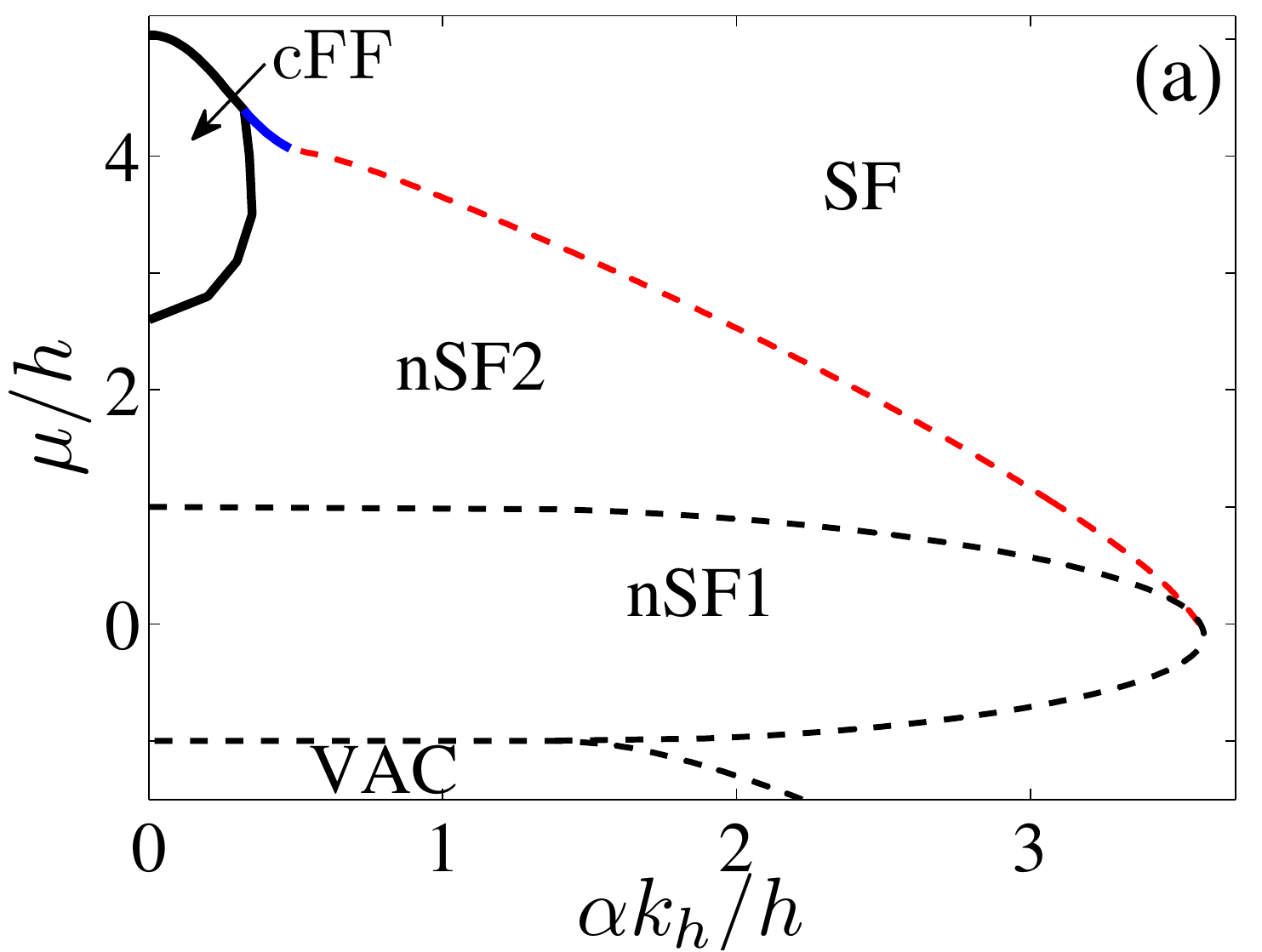}
\includegraphics[width=5cm]{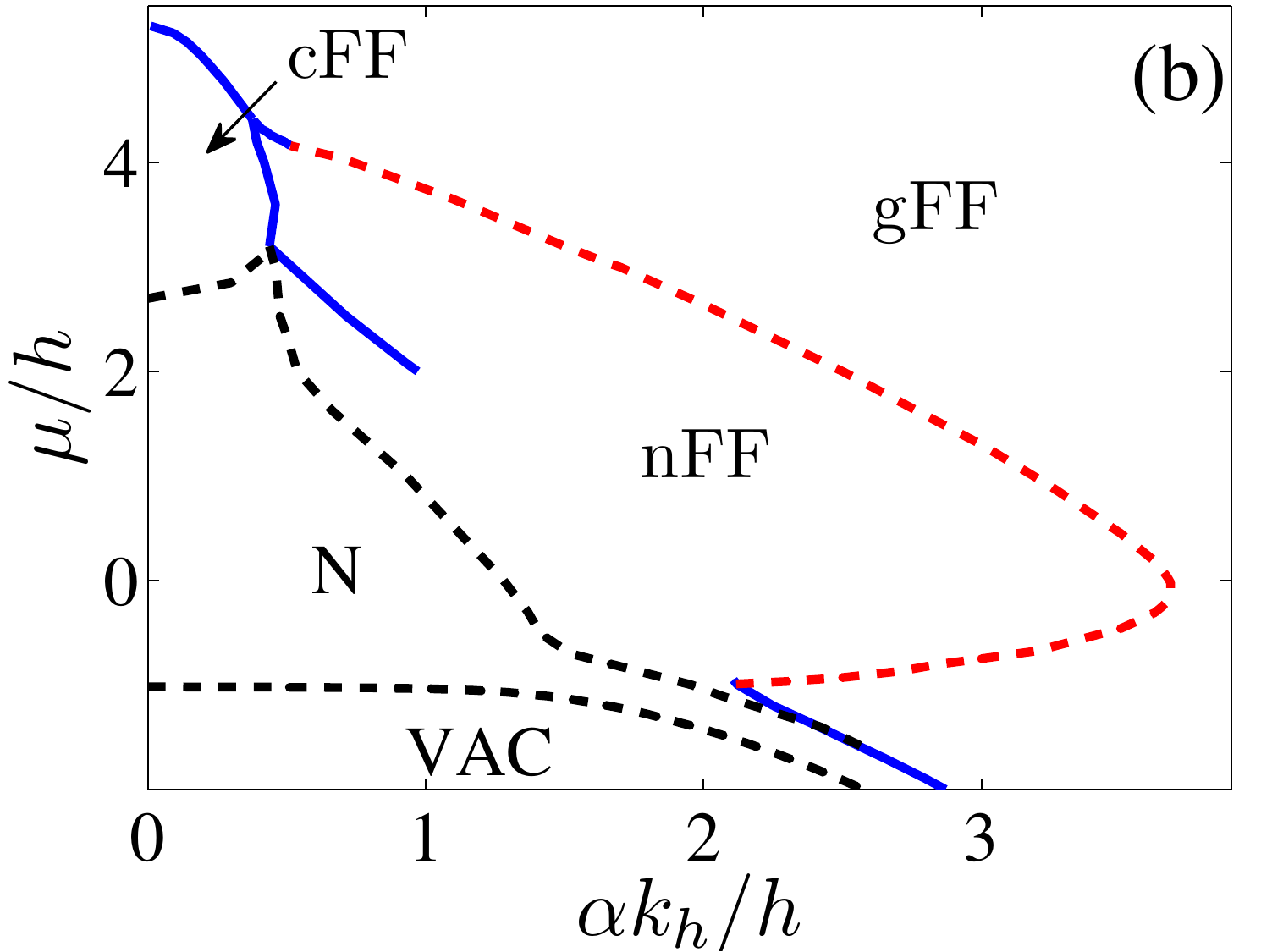}
\includegraphics[width=5cm]{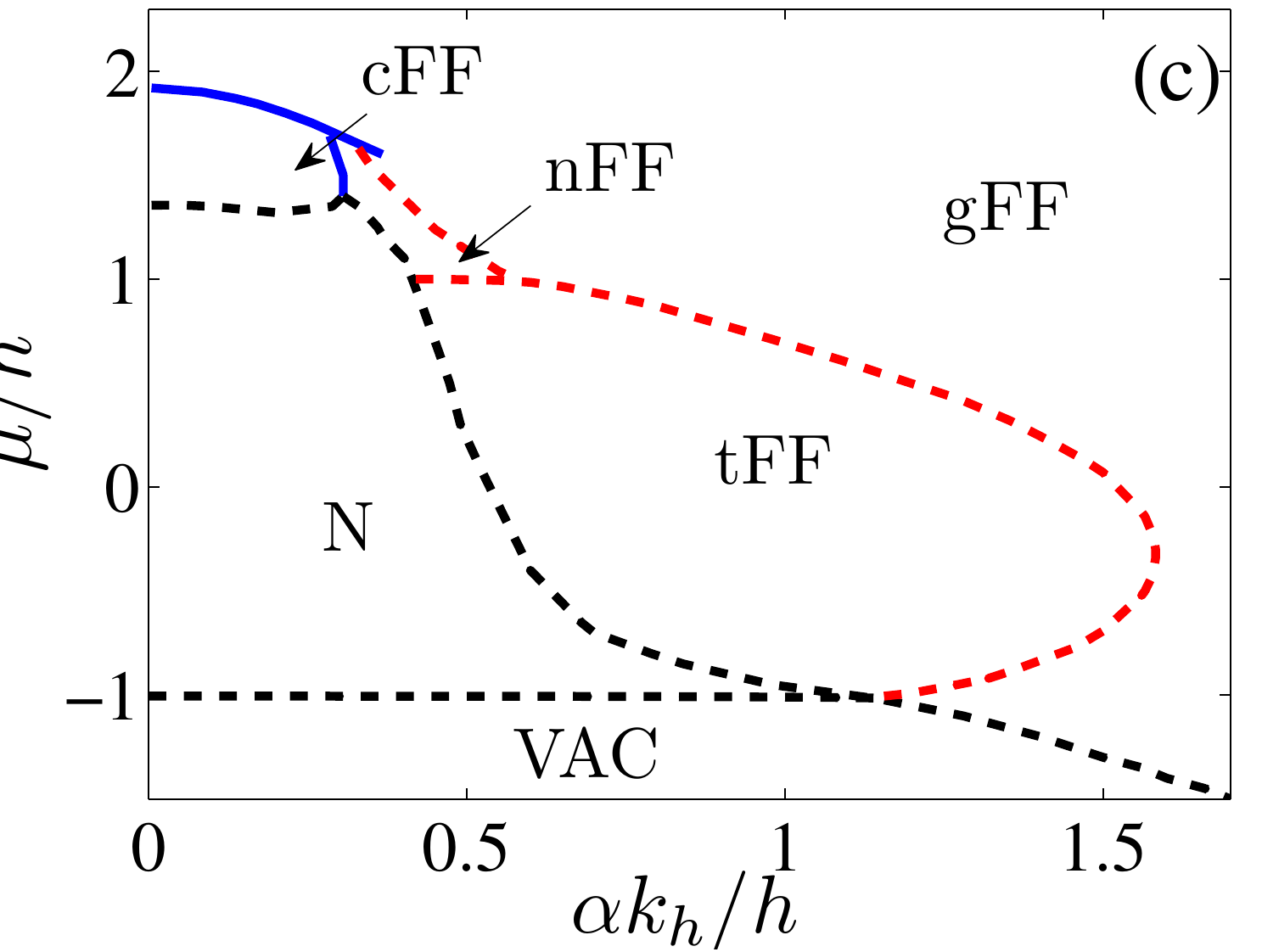}
\caption{ Typical phase diagrams for a two-dimensional Fermi gas under: (a) NIST SOC without two-photon detuning (no in-plane Zeeman field); (b) NIST SOC with finite two-photon detuning (finite in-plane Zeeman field); (c) Rashba SOC with both the out-of-plane and in-plane Zeeman field. The solid curves are first-order boundaries, the dashed curves represent continuous phase boundaries. Here, VAC represents the vacuum, and N represents normal state. See Ref.~\cite{wyfflo,tfflo2}}\label{fig_socchapffphase}
\end{figure*}

In the context of ultracold atomic gases, a natural question is whether the topological superfluid state and the associated Majorana zero modes can be prepared and probed. Although the experimental realizaiton of the Rashba SOC has not been achieved yet, there have been various theoretical proposals on its implementation in cold atoms, which gives rise to even more theoretical characterizations of the topological superfluid phase and the related phase transitions in the context of ultracold Fermi gases. In two dimensions, it is found that a topological superfluid phase which supports Majorana zero modes can be stabilized~\cite{zhang,sato}. In a uniform system, these different phases will give rise to phase-separated states with various first-order boundaries (see Fig.~\ref{fig_Eb}). In a trapping potential, which is more relevant to the experimental conditions, these different phases naturally separate in space, leading to shell structures similar to those in a polarized Fermi gas~\cite{wy2d}.

Another intriguing finding on the mean-field phase diagram Fig.~\ref{fig_Eb} is the absence of a normal phase, even in the limit of large magnetic field $\alpha k_F / h \to 0$. Instead of being zero in the case without SOC, here, the order parameter decreases exponentially with increasing $h$, but remains finite within the entire SF and TSF regime. This somewhat counterintuitive result is an artifact of the mean-field approach, which completely neglects quantum and thermal fluctuations. A more careful analysis incorporating particle-hole fluctuations in the large-Zeeman-field limit shows that there exits a so-called polaron-molecule transition, suggesting that the true ground state is a normal phase, at least in the weak-coupling regime~\cite{polaronwy}.

When the Rashba SOC is replaced by the less symmetric NIST SOC, the topological superfluid phase would be replaced by various nodal superfluid phases (nSF1 and nSF2), with either two or four discrete nodal points along the direction with no SOC. The mean-field zero-temperature phase diagram in two dimensions is shown in Fig.~\ref{fig_socchapffphase}(a). Note that the cFF phase is due to the interbranch pairing mechanism and not the SOC-induced FF states that we have discussed previously. In three dimensions, neither the Rashba SOC nor the NIST SOC would lead to a topological superfluid, as the pairing superfluid phases therein are either fully gapped trivial superfluid (SF) state or nodal superfluid phase. Interestingly, for gapless superfluid states in a three dimensional Fermi gas with the NIST SOC, the nodal points typically form closed surfaces in momentum space~\cite{iskinnistsoc}.

\subsection{SOC-induced FF pairing and topological FF state}

An important consequence of the intrabranch pairing is the dramatically enhanced FF pairing states under SOC and Fermi surface asymmetry. As illustrated in Fig.~\ref{fig_bcspairing}(c), under an additional in-plane Zeeman field, the Fermi surface is deformed,
such that it no longer has inversion symmetry along the axis of the transverse field. In the weakly interacting limit, it is clear that a simple BCS pairing state with zero center-of-mass momentum becomes energetically unfavorable. This opens up the possibility of an exotic FFLO-like pairing state with finite center-of-mass momentum~\cite{wyfflo}. From the general argument above, we may further infer that such a pairing state is a natural result of the co-existence of SOC and Fermi surface asymmetry, and should generally exist in such systems, regardless of the exact type of SOC in the system. Indeed, these exotic pairing states have been reported to exist in various systems of different dimensions and with different forms of SOC~\cite{puhan3dsoc,xiangfa3dsoc,tfflo0,tfflo1,tfflo2,chuanwei3dsoc,tfflohu,huhuintFF,huhui3dsoc,iskinnistsoc,patrick,zhengzhenrashba}.

To illustrate this, we first consider an experimentally relevant system, where the NIST SOC is imposed on a two-dimensional Fermi gas with effective axial and transverse Zeeman fields. Similar to the Rashba case, a mean-field description of the pairing states here can be seen as a natural extension of the standard BCS theory
\begin{eqnarray}
H_{\text{eff}}&=&\frac{1}{2}\sum_{\mathbf{k}}\begin{pmatrix}
\lambda^{+}_{\cp k}&0&h&\Delta_{\cp Q}\\
0&-\lambda^{+}_{{\cp Q}-{\cp k}}&-\Delta^{\ast}_{\cp Q}&-h\\
h&-\Delta_{\cp Q}&\lambda^{-}_{\cp k}&0\\
\Delta^{\ast}_{\cp Q}&-h&0&-\lambda^{-}_{{\cp Q}-{\cp k}}
\end{pmatrix}\nonumber\\
&+&\sum_{\mathbf{k}}\xi_{|\mathbf{Q}-\mathbf{k}|}-\frac{|\Delta_{\cp Q}|^2}{U},\label{eqnHeffsoc}
\end{eqnarray}
where $\lambda_{\cp k}^{\pm}=\xi_{\cp k}\pm\alpha k_x\mp h_x$, the order parameter $\Delta_{\cp Q}=U\sum_{\mathbf{k}}\left\langle a_{\mathbf{Q}-\mathbf{k}\downarrow} a_{\mathbf{k}\uparrow} \right\rangle$. The Hamiltonian (\ref{eqnHeffsoc}) has been written under the hyperfine-spin basis $\left\{a_{\mathbf{k}\uparrow},a^{\dag}_{\mathbf{Q}-\mathbf{k}\uparrow},a_{\mathbf{k}\downarrow},a^{\dag}_{\mathbf{Q}-\mathbf{k}\downarrow}\right\}^{T}$. Again, the SOC parameter $\alpha$ is related to the momentum transfer of the Raman process in the NIST scheme, and the effective Zeeman field $h$ and $h_x$ are proportional to the effecitve Rabi-frequency and the two-photon detuning of the Raman lasers, respectively. The zero-temperature thermodynamic potential can be obtained by diagonalizing the effective Hamiltonian
\begin{equation}
\Omega=\sum_{\mathbf{k}}\xi_{|\mathbf{Q}-\mathbf{k}|}+\sum_{\mathbf{k},\nu}\theta(-E^{\eta}_{\mathbf{k},\nu})E^{\eta}_{\mathbf{k},\nu}-\frac{|\Delta_{\cp Q}|^2}{U},
\label{eqnOmegasoc}
\end{equation}
where the quasiparticle (quasihole) dispersion $E^{\eta}_{\mathbf{k},\nu}$ ($\nu=1,2$, $\eta=\pm$) are the eigenvalues of the matrix in Hamiltonian (\ref{eqnHeffsoc}), and $\theta(x)$ is the Heaviside step function.

From the previous general analysis in the weak-coupling limit, the BCS pairing states with zero center-of-mass momentum would become unstable against an FF pairing state under the Fermi surface asymmetry. This implies an instability of the BCS state with finite $h_x$. This point can be demonstrated by performing a small $\mathbf{Q}$ expansion around the local minimum in the thermodynamic potential landscape that corresponds to the BCS pairing state
\begin{equation}
\Omega(\Delta,Q_x)=\Omega_0(\Delta)+\Omega_1(\Delta)Q_x+\Omega_2(\Delta)Q_x^2+{\cal O}(Q_x^3),
\end{equation}
where we have assumed $\mathbf{Q}=(Q_x,0)$. It is then straightforward to demonstrate numerically that for $h_x=0$, we have $\Omega_1=0$, $\Omega_2>0$; while for $h_x\neq 0$, we have $\Omega_1\neq 0$, which has an opposite sign to that of $h_x$. This is a direct evidence that the BCS pairing state in the presence of Fermi surface asymmetry and SOC becomes unstable against an FF state, with the center-of-mass momentum $\mathbf{Q}$ opposite to the direction of the transverse field $h_x$. A qualitative understanding of these FF states is that the combination of SOC and Fermi surface asymmetry shifts the local minima that corresponds to BCS pairing states ($Q=0$) onto the finite-$\mathbf{Q}$ plane. This is further reflected by the observation that the magnitude of the center-of-mass momentum $Q$ decreases as Fermi surface asymmetry becomes smaller, i.e., with increasing chemical potential $\mu$ or SOC strength $\alpha$~\cite{wyfflo,wyfflolong}.

We show the mean-field phase diagram in Fig.~\ref{fig_socchapffphase}(b). Recalling Fig.~\ref{fig_Eb}, it is clear that the topological superfluid state under the Rashba SOC is now replaced by a nodal FF (nFF) state, while the trivial superfluid state is replaced by a gapped FF (gFF) state. These SOC-induced FF states originate from intrabranch pairing, and are different in mechanism from the conventional FF state.

Among these SOC-induced FF states, perhaps the most interesting case is the topological FF (tFF) state, where the pairing state can have nonzero center-of-mass momentum and topologically nontrivial properties simultaneously~\cite{tfflo0,tfflo1,tfflo2,tfflohu}. This exotic pairing phase can be understood as derived from a typical topological superfluid phase in two dimensions, where Rashba SOC, $s$-wave pairing order and an out-of-plane Zeeman field coexist. With the addition of another effective Zeeman field in the transverse direction, the Fermi surface becomes asymmetric, and according to the preceding analysis, the ground state of the system necessarily acquires a nonzero center-of-mass momentum. More importantly, the ground state would inherit all topological properties from the topological superfluid state, provided that the deformation of the Fermi surface should not be drastic enough to close the bulk gap. The phase diagram of a two-dimensional Fermi gas under Rashba SOC and cross Zeeman fields is shown in Fig.~\ref{fig_socchapffphase}(c). Apparently, the stability region of the tFF state roughly corresponds to the topological superfluid state in Fig.~\ref{fig_Eb}.

\section{Engineering novel states}\label{sec_engineer}

An advantage of ultracold atomic gases is their highly tunable parameters. While SOC provides the possibility of engineering single-particle dispersion, in principle, one can also change the interaction strength via the well-established Feshbach resonance technique. In fact, when combining SOC with other available tools in ultracold atomic gases, we can engineer highly nontrivial states.

\subsection{Engineering FF state}\label{sec51}

\begin{figure}
\centering
\includegraphics[width=7cm]{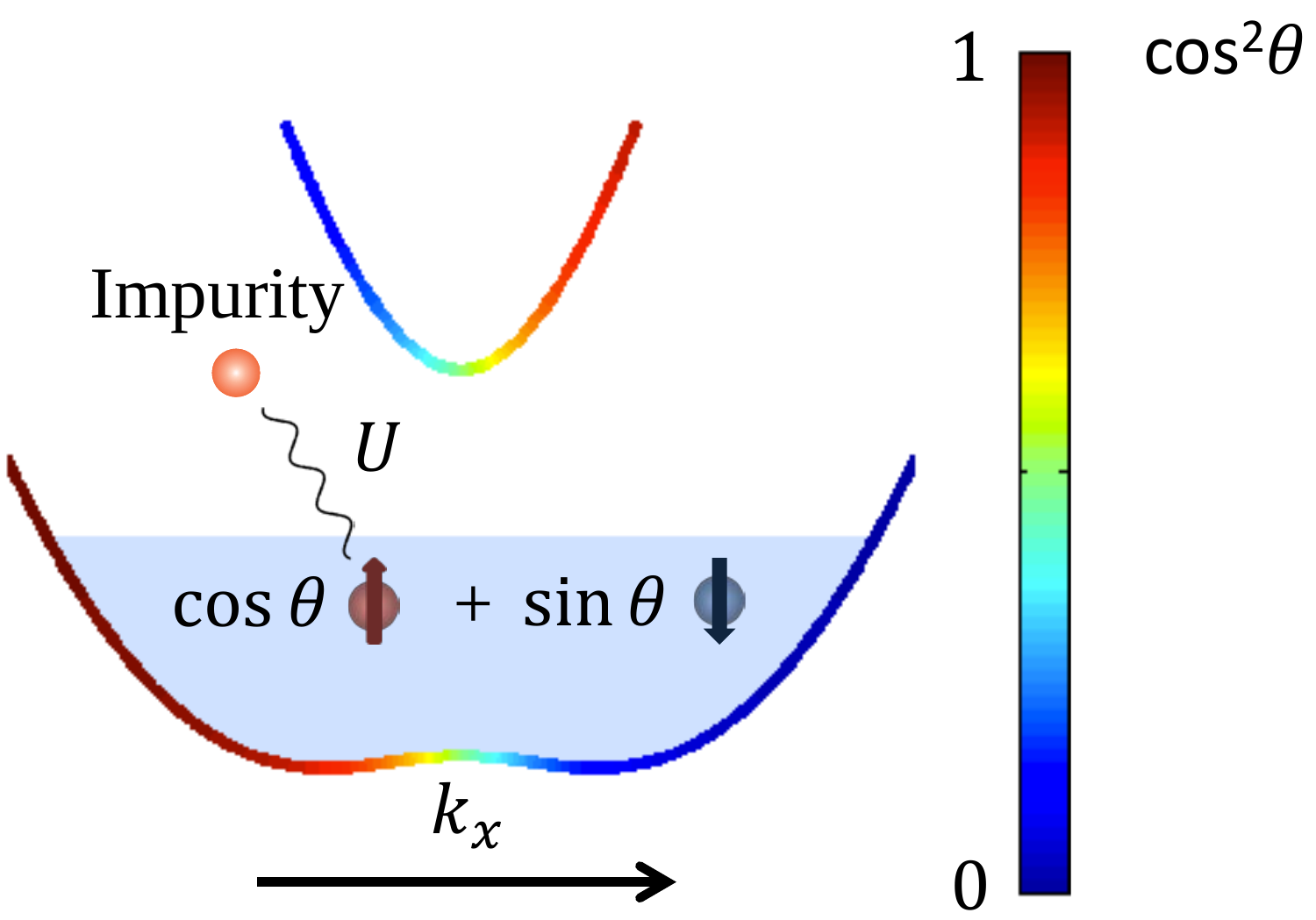}
\caption{Schematics of the three-component Fermi system in Sec.~\ref{sec51}. The impurity atoms interact only with the spin-up atoms in a spin-orbit coupled Fermi gas.
The fractions of spin components in both helicity branches are momentum dependent, as characterized by $\cos\theta=-\beta_{\cp k}^{-}$ and  $\sin\theta=\beta_{\cp k}^{+}$, where $\beta_{\cp k}^{\pm}$ is defined in Sec.~\ref{sec_singlespec}. See Ref.~\cite{cuiyipairing}.}
\label{fig_schematic}
\end{figure}

\begin{figure}
\centering
\includegraphics[width=9cm]{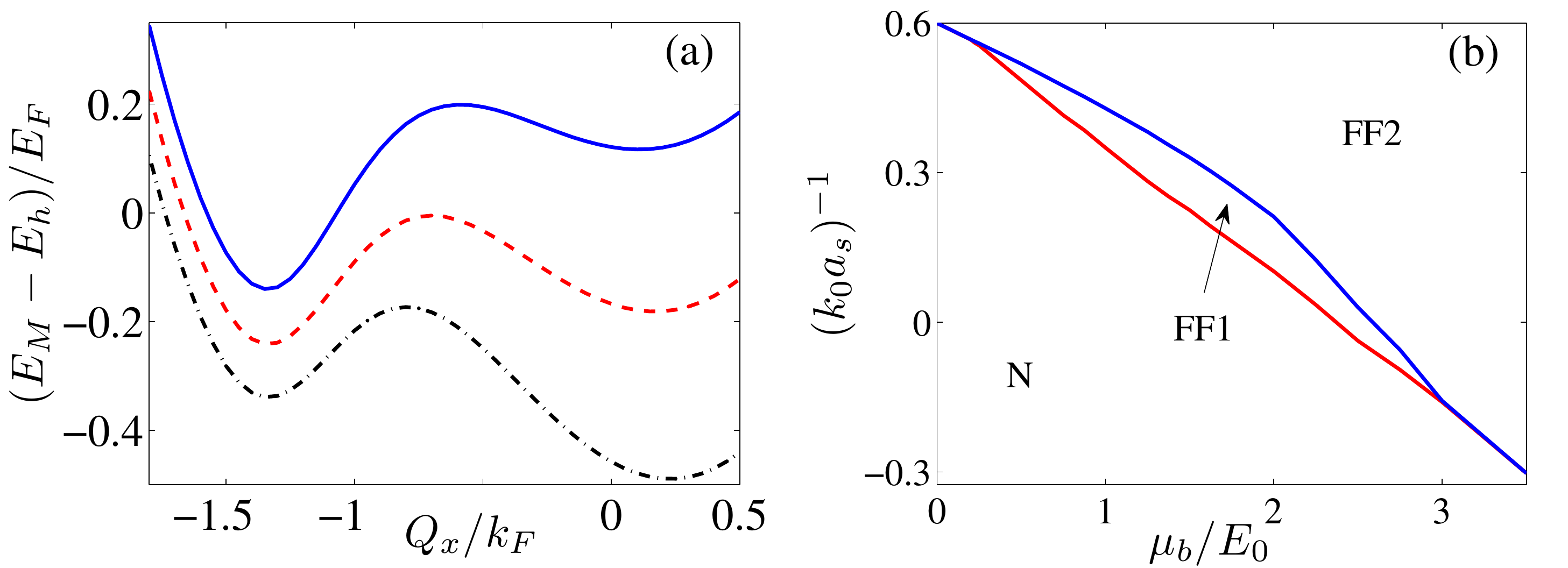}
\caption{(a) Competition between molecules of different center-of-mass momenta as the actual Fermi energy changes. Molecular energies $E_M$ relative to the Fermi energy $E_h$ as functions of the center-of-mass momentum $Q_x$, with $h/E_F=1.2$ (solid), $h/E_F=1$ (dashed), and $h/E_F=0.8$ (dash-dotted). The unit of energy is the Fermi energy $E_F$ of a two-component, noninteracting Fermi gas in the absence of SOC and with the same total number density. (b) Typical phase diagram in the ($\mu_b/E_0, (k_0a_s)^{-1}$) plane in a Li-K-K system. The red (blue) curve shows the transition boundary between the normal state (N) and FF states with different center-of-mass momentum (FF1 and FF2). See Ref.~\cite{cuiyipairing}.}
\label{fig_compete}
\end{figure}

As an example, we consider a three-component Fermi gas, where one fermion species (`impurity') is tuned close to a wide Feshbach resonance with one of the spin-species in a two-component Fermi gas under the NIST SOC~\cite{cuiyipairing}. The pairing mechanism here is illustrated in Fig.~\ref{fig_schematic}. Under the NIST SOC, the spin components in both helicity branches are momentum dependent. The pairing states naturally acquire a nonzero center-of-mass momentum, which is dependent on the position of the Fermi surface as well as the interaction strength. For instance, in the weak-coupling limit, when the Fermi surface lies in the lower branch, the center-of-mass momentum is negative, since the pairing is lower-branch dominated; while if the Fermi surface lies in the upper branch, the center-of-mass momentum can be positive, as the ground state is the result of the competition between lower-branch-dominated pairing and the upper-branch-dominated pairing. This can lead to the competition between pairing states with different center-of-mass momentum, as illustrated in Fig.~\ref{fig_compete}. Apparently, the FF pairing states in this system is different from the SOC-induced FF states discussed previously. Qualitatively, we are replacing the Fermi surface asymmetry in the previous section with the asymmetry in the spin degrees of freedom of the helicity basis.

\begin{figure}
\centering
\includegraphics[width=7cm]{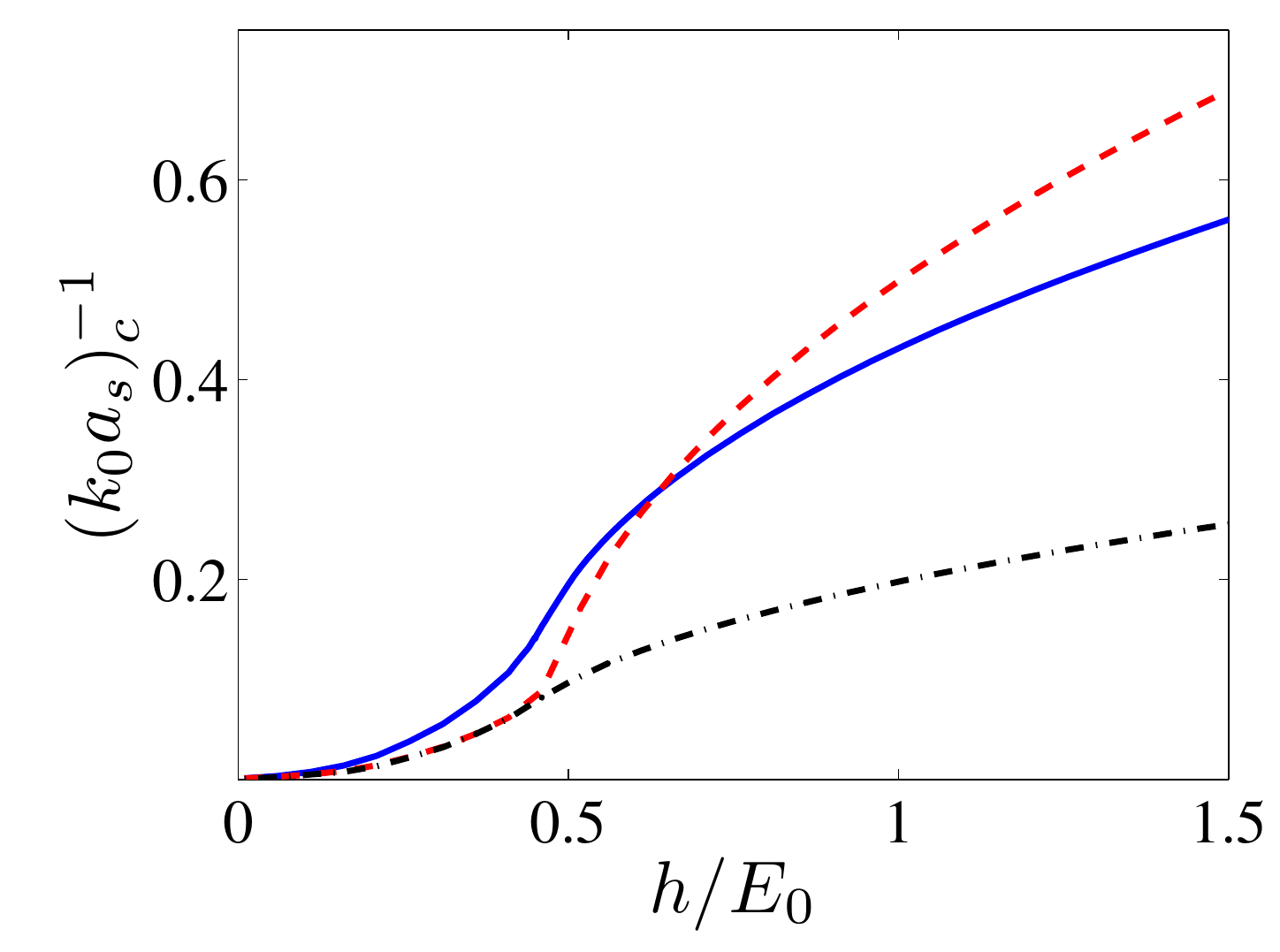}
\caption{ Two-body bound state threshold for varying SOC parameter $h$ and a fixed $\alpha k_0/E_0=1$. The mass ratio $\eta=m_a/m_b$, where $m_b$ is the mass of the impurity atom, $m_a$ is the mass of the atoms under SOC. Here, $\eta=1$ (solid), $\eta=6/40$ (dashed), and $\eta=40/6$ (dash-dotted). The unit of energy $E_0=2m_a\alpha^2/\hbar^2$, and the unit of momentum $k_0=2m_a\alpha/\hbar^2$. See Ref.~\cite{cuiyipairing}.}
\label{fig_twobodythresh}
\end{figure}

A notable feature of the current system is that the two-body bound state is suppressed by SOC. In the absence of SOC, the spin-down atoms are decoupled, and the two-body bound state between the `impurity' and the spin-up atom emerges at resonance with $a_s^{-1}=0$, where $a_s$ is the scattering length between the spin-up atom and the `impurity' atom. Under SOC, this two-body bound state threshold is pushed toward the BEC limit with $a_s^{-1}>0$ (see Fig.~\ref{fig_twobodythresh}). Roughly, as SOC establishes `correlation' between the spin-up and the spin-down atoms, it is more difficult for the `impurity' and the spin-up atoms to combine and form a bound state. As we will see later, a more rigorous description of this peculiar `correlation' is the symmetry in the SOC-modified single-particle dispersion. This suppression of two-body bound state naturally leads to the question of the stability of three-body bound states, i.e., whether a three-body bound state can be more stable than the two-body bound state in certain parameter region?

\subsection{Exotic trimer states under SOC}\label{sec52}

\begin{figure}[tbp]
\centering
\includegraphics[width=7cm]{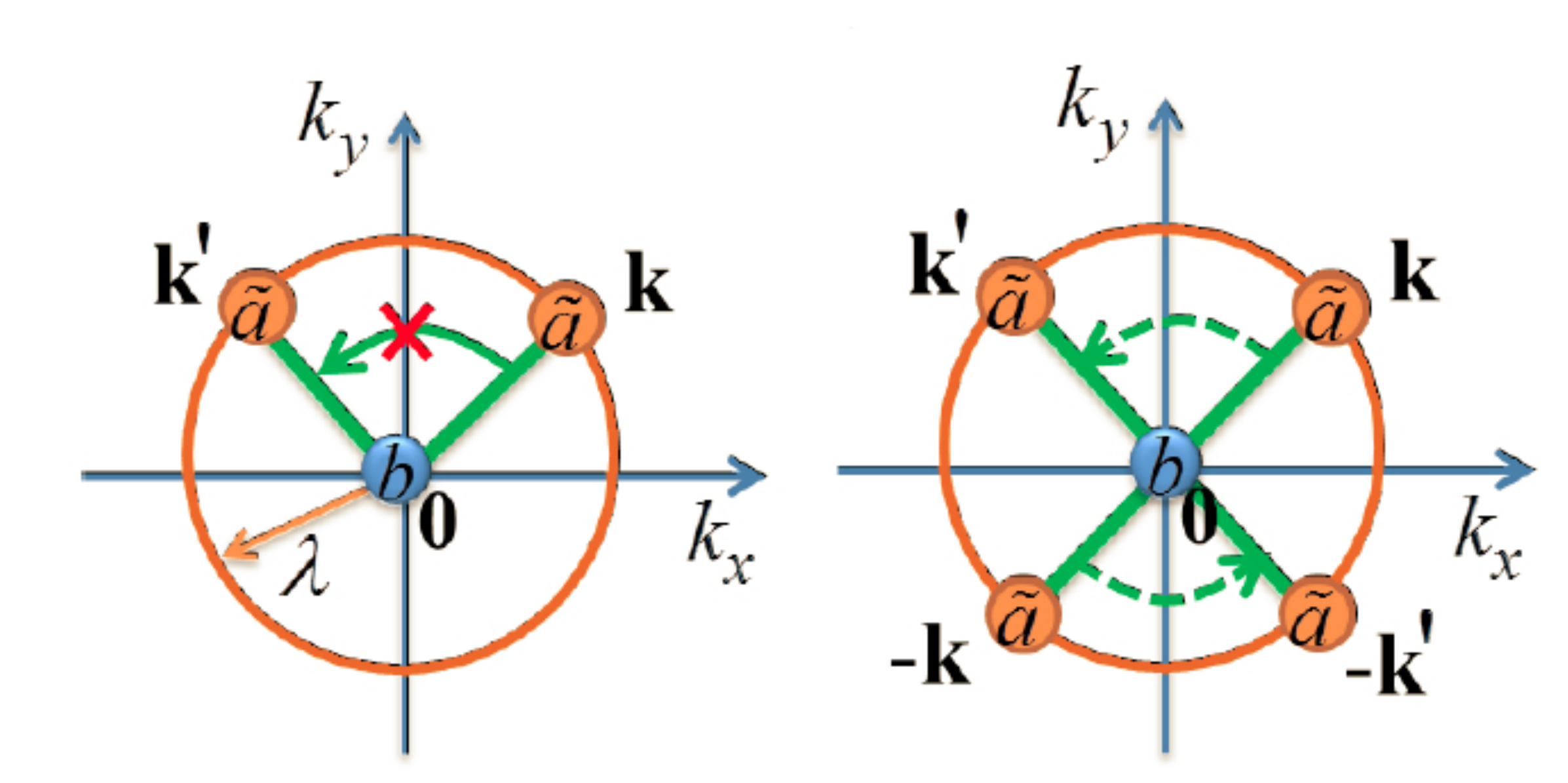}
\includegraphics[width=7cm]{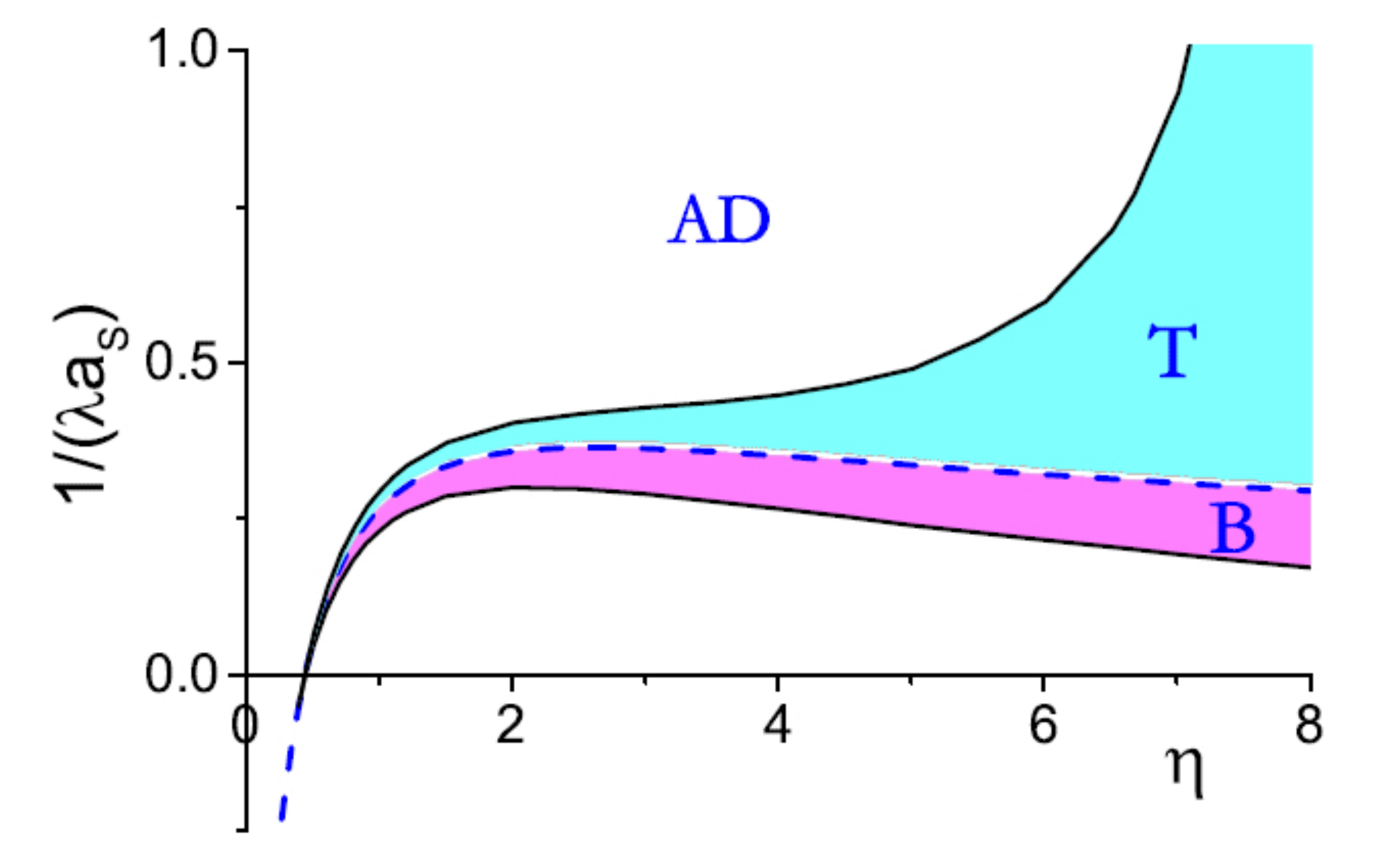}
\caption{ (Upper panel) Illustration of the Borromean binding mechanism in the
the three-component Fermi system in Sec.~\ref{sec52}. In this $\tilde{a}-\tilde{a}-b$ system, $\tilde{a}$ atoms are subject to the Rashba SOC, and $b$ serves as the `impurity'. (Upper left) The two-body scattering within the lowest-energy subspace of the $\tilde{a}-b$ system cannot proceed due to the conservation of total momentum.
(Upper right) The three-body scattering within the lowest energy subspace of the $\tilde{a}-\tilde{a}-b$ system is allowed (green dashed arrows). (Lower panel) Phase diagram for the system in Sec.~\ref{sec52} on the $(1/(\lambda a_s),\eta)$ plane, where $\lambda=m_a\alpha/\hbar^2$. The lower and upper solid curves respectively represent the threshold of Borromean (`B') state and the boundary at which the ordinary trimer (`T') merges into the atom-dimer continuum (`AD'). The blue dashed curve is the dimer threshold, which is also the boundary between `B' and `T' for $\eta\geqslant 0.39$. See Ref.~\cite{trimer2}}
\label{fig_borromean}
\end{figure}

A short answer to the question above is: not with the NIST SOC~\cite{cuiyipairing}. However, it is possible to stabilize three-body bound states, or trimers, if we consider other forms of SOC, especially those with a high symmetry. Indeed, if we replace the NIST SOC with the Rashba SOC in the previously discussed three-component Fermi system, trimers can be stabilized even in the absence of any two-body bound states~\cite{trimer2}. The stabilization of this so-called Borromean state can be understood from the special symmetry of the Rashba-modified single-particle dispersion spectra (see Fig.~\ref{fig_borromean}). As illustrated in Fig.~\ref{fig_specsoc}(c), under the Rashba SOC, the degenerate subspace of the single-particle ground state forms a ring in momentum space. With such a spectral symmetry, the scattering within the lowest-energy subspace is blocked due to the total momentum conservation (see Fig.~\ref{fig_borromean}), which effectively suppresses the formation of a two-body bound state. In contrast, the three-body scattering is not blocked. Under the NIST SOC, the single-particle dispersion spectra have less symmetry. Hence, although the two-body bound states are also suppressed in those systems, they remain to be energetically more favorable than the trimers.

Borromean states have been found before in different physical systems. However, previously studied Borromean states are nonuniversal, with properties dependent on the short-range details of the two-body interaction potential~\cite{Efimov,halo1,Braaten}. Here, the Borromean states are universal, with properties only dependent on the two-body $s$-wave scattering length and the SOC coupling strength. Furthermore, near the trimer threshold of such a system, three-body resonances should give rise to interesting many-body states with exotic few-body correlations. The universal Borromean state in the spin-orbit coupled Fermi system is thus highly nontrivial. Note that universal trimers have also been identified in a related system, where an atom under isotropic SOC interacts with two other atoms~\cite{trimer1}. Both of these cases suggest that exotic few-body states may be stabilized by introducing SOC as another dimension of control.

\section{Summary}\label{sec_fin}

In this brief review, we have focused on the general pairing mechanism in a spin-orbit coupled Fermi gas. As SOC changes the single-particle dispersion spectra, pairing superfluidity in the system takes on many new and exotic forms: topological superfluid, SOC-induced FF state, topological superfluid so on so forth. Besides these exotic pairing states, SOC can also lead to novel universal trimer states, which features exotic few-body correlations upon which even more interesting many-body phases may emerge. In all these cases, by modifying the single-particle dispersion, SOC proves to be a powerful tool of quantum control, which, when combined with the outstanding tunability of ultracold atomic gases, is playing an increasingly important role in fulfilling the potential of quantum simulation with ultracold atomic gases.

\acknowledgments
This work is supported by NFRP (2011CB921200, 2011CBA00200), NKBRP (2013CB922000), NNSF (60921091), NSFC (10904172, 11104158,11374177,11105134, 1127409,11374283), the Fundamental Research Funds for the Central Universities (WK2470000006), the Research Funds of Renmin University of China (10XNL016), and the programs of Chinese Academy of Sciences.


\begin{thebibliography}{99}
\bibitem{gauge1exp} Y.-J. Lin, R. L. Compton, A. R. Perry, W. D. Phillips, J. V. Porto, and I. B. Spielman, Phys. Rev. Lett. {\bf 102}, 130401 (2009).

\bibitem{gaugenist1} Y.-J. Lin, R. L. Compton, K. Jim\'{e}nez-Garc\'{i}a, W. D. Phillips, J. V. Porto, I. B. Spielman, Nature {\bf 462}, 628 (2009).

\bibitem{gaugenist2} Y.-J. Lin, R. L. Compton, K. Jim\'{e}nez-Garc\'{i}a, W. D. Phillips, J. V. Porto, and I. B. Spielman, Nat. Phys. {\bf 7}, 531 (2011).

\bibitem{gauge2exp} Y.-J. Lin, K. Jim\'{e}nez-Garc\'{i}a, and I. B. Spielman, Nature (London) {\bf 471}, 83 (2011).

\bibitem{fermisocexp1} P. Wang, Z.-Q. Yu, Z. Fu, J. Miao, L. Huang, S. Chai, H. Zhai, and J. Zhang, Phys. Rev. Lett. {\bf 109}, 095301 (2012).

\bibitem{fermisocexp2} L. W. Cheuk, A. T. Sommer, Z. Hadzibabic, T. Yefsah, W. S. Bakr, and M. W. Zwierlein, Phys. Rev. Lett. {\bf 109}, 095302 (2012).

\bibitem{shuaiexp} J.-Y. Zhang, S.-C. Ji, Z. Chen, L. Zhang, Z.-D. Du, B. Yan, G.-S. Pan, B. Zhao, Y. Deng, H. Zhai, S. Chen, and J.-W. Pan, Phys. Rev. Lett. {\bf 109}, 115301 (2012).

\bibitem{engelsexp} C. Qu, C. Hamner, M. Gong, C. Zhang, and P. Engels, Phys. Rev. A {\bf 88}, 021604(R) (2013).

\bibitem{nistfesh} R. A. Williams, M. C. Beeler, L. J. LeBlanc, K. Jimenez-Garcia, and I. B. Spielman, Phys. Rev. Lett. {\bf 111}, 095301 (2013).

\bibitem{zjnatphys} Z. Fu, L. Huang, Z. Meng, P. Wang, L. Zhang, S. Zhu, H. Zhai, P. Zhang, and J. Zhang, Nat. Phys. {\bf 10}, 110 (2014).

\bibitem{shuainatphys} S.-C. Ji, J.-Y. Zhang, L. Zhang, Z.-D. Du, W. Zheng, Y.-J. Deng, H. Zhai, S. Chen, and J.-W. Pan, Nat. Phys. {\bf 10}, 314 (2014).

\bibitem{chenyong} A. J. Olson, S.-J. Wang, R. J. Niffenegger, C.-H. Li, C. H. Greene, and Y. P. Chen, Phys. Rev. A {\bf 90}, 013616 (2014).

\bibitem{kanereview} M. Z. Hasan and C. L. Kane, Rev. Mod. Phys. {\bf 82}, 3045 (2010).

\bibitem{zhangscreview} X.-L. Qi and S.-C. Zhang, Rev. Mod. Phys. {\bf 83}, 1057 (2011).

\bibitem{alicea} J. Alicea, Rep. Prog. Phys. {\bf 75}, 076501 (2012).

\bibitem{Rashba_Spielman_1}D. L. Campbell, G. Juzeli\={u}nas, and I. B. Spielman, Phys. Rev. A {\bf 84}, 025602 (2011).

\bibitem{Rashba_Spielman_2}J. D. Sau, R. Sensarma, S. Powell, I. B. Spielman, and S. Das Sarma, Phys. Rev. B {\bf 83}, 140510(R) (2011).

\bibitem{Rashba_Xu_1} Z. F. Xu and L. You, Phys. Rev. A {\bf 85}, 043605 (2012).

\bibitem{Rashba_Liu} X.-J. Liu, K. T. Law, and T. K. Ng, Phys. Rev. Lett. {\bf 112}, 086401 (2014).

\bibitem{Rashba_Spielman_3}B. M. Anderson, I. B. Spielman, and G. Juzeli\={u}nas, Phys. Rev. Lett. {\bf 111}, 125301 (2013).

\bibitem{Rashba_Xu_2}Z.-F. Xu, L. You, and M. Ueda, Phys. Rev. A {\bf 87}, 063634 (2013)

\bibitem{congjun3d}  Y. Li, X. Zhou, and C. Wu, Phys. Rev. B {\bf 85}, 125122 (2012).

\bibitem{spielman_3d_1} B. M. Anderson, G. Juzeli\={u}nas, V. M. Galitski and I. B. Spielman, Phys. Rev. Lett. {\bf 108}, 235301 (2012).

\bibitem{zhang} C. Zhang, S. Tewari, R. M. Lutchyn, and S. Das Sarma, Phys. Rev. Lett. {\bf 101}, 160401 (2008).

\bibitem{sato} M. Sato, Y. Takahashi, and S. Fujimoto, Phys. Rev. Lett. {\bf 103}, 020401 (2009).

\bibitem{gongzhang} M. Gong, S. Tewari, and C. Zhang, Phys. Rev. Lett. {\bf 107}, 195303 (2011).

\bibitem{2d2} L. Dell'Anna, G. Mazzarella, and L. Salasnich, Phys. Rev. A {\bf 84}, 033633 (2011).

\bibitem{wy2d} J. Zhou, W. Zhang, W. Yi, Phys. Rev. A {\bf 84}, 063603 (2011).

\bibitem{2d1} M. Gong, G. Chen, S. Jia, and C. Zhang, Phys. Rev. Lett. {\bf 109}, 105302 (2012).

\bibitem{wmliu} R. Liao, Y. Yi-Xiang, and W.-M. Liu, Phys. Rev. Lett. {\bf 108}, 080406 (2012).

\bibitem{helianyi} L. He and X.-G. Huang, Phys. Rev. Lett. {\bf 108}, 145302 (2012).

\bibitem{tfflo0} C. Chen, Phys. Rev. Lett. {\bf 111}, 235302 (2013).

\bibitem{tfflo1} C. Qu, Z. Zheng, M. Gong, Y. Xu, L. Mao, X. Zou, G. Guo, and C. Zhang, Nat. Commun. {\bf 4}, 2710 (2013).

\bibitem{tfflo2} W. Zhang and W. Yi, Nat. Commun. {\bf 4}, 2711 (2013).

\bibitem{tfflohu} X.-J. Liu and H. Hu, Phys. Rev. A {\bf 88}, 023622(R) (2013).

\bibitem{chuanwei3dsoc} Y. Xu, R.-L. Chu and C. Zhang, Phys. Rev. Lett. {\bf 112}, 136402 (2014).

\bibitem{huhuintFF} Y. Cao, S.-H. Zou, X.-J. Liu, G.-L. Long and H. Hu, arXiv:1402.6832.

\bibitem{huhui3dsoc} H. Hu, L. Dong, Y. Cao, H. Pu, and X.-J. Liu, arXiv:1404.2442.

\bibitem{iskin} M. Iskin and A. L. Subasi, Phys. Rev. Lett. {\bf 107}, 050402 (2011).

\bibitem{thermo} W. Yi and G.-C. Guo, Phys. Rev. A {\bf 84}, 031608(R) (2011).

\bibitem{melo} L. Han and C. A. R. S\'{a} de Melo, Phys. Rev. A {\bf 85} 011606(R) (2012).

\bibitem{polaronwy} W. Yi and W. Zhang, Phys. Rev. Lett. {\bf 109}, 140402 (2012).

\bibitem{xiaosen} X. Yang and S. Wan, Phys. Rev. A {\bf 85}, 023633 (2012).

\bibitem{iskinnistsoc} M. Iskin and A. L. Subasi, Phys. Rev. A {\bf 87}, 063627 (2013).

\bibitem{puhan3dsoc} L. Dong, L. Jiang, and H. Pu, New J. Phys. {\bf 15}, 075014 (2013).

\bibitem{xiangfa3dsoc} X.-F. Zhou, G.-C. Guo, W. Zhang, and W. Yi, Phys. Rev. A {\bf 87}, 063606 (2013).

\bibitem{wyfflo} F. Wu, G.-C. Guo, W. Zhang, and W. Yi, Phys. Rev. Lett. {\bf 110}, 110401 (2013).

\bibitem{wyfflolong} F. Wu, G.-C. Guo, W. Zhang, and W. Yi, Phys. Rev. A {\bf 88}, 043614 (2013).

\bibitem{cuiyipairing} L. Zhou, X. Cui, and W. Yi, Phys. Rev. Lett. {\bf 112}, 195301 (2014).

\bibitem{trimer1} Z.-Y. Shi, X. Cui, and H. Zhai, Phys. Rev. Lett. {\bf112}, 013201 (2014).

\bibitem{trimer2} X. Cui and W. Yi, Phys. Rev. X {\bf 4}, 031026 (2014).

\bibitem{hzsocbec} C. Wang, C. Gao, C. Jian, and H. Zhai, Phys. Rev. Lett. {\bf 105}, 160403 (2010).

\bibitem{cplwu} C. Wu, I. Mondragon-Shem, and X.-F. Zhou, Chin. Phys. Lett. {\bf 28} 097102 (2011).

\bibitem{soc3} J. P. Vyasanakere, S. Zhang, and V. B. Shenoy, Phys. Rev. B {\bf 84}, 014512 (2011).

\bibitem{soc4} Z.-Q. Yu and H. Zhai, Phys. Rev. Lett. {\bf 107}, 195305 (2011).

\bibitem{dalibardreview} J. Dalibard, F. Gerbier, G. Juzeli\={u}nas, and P. \"{O}hberg, Rev. Mod. Phys. {\bf 83}, 1523 (2011).

\bibitem{xjlgauge} X.-J. Liu, M. F. Borunda, X. Liu, and J. Sinova, Phys. Rev. Lett. {\bf 102}, 046402 (2009).

\bibitem{nistsoctheory} I. B. Spielman, Phys. Rev. A {\bf 79}, 063613 (2009).

\bibitem{congjun3dsoc} Y. Li, X. Zhou, and C. Wu, arXiv:1205.2162.

\bibitem{ngoldman1} N. Goldman, F. Gerbier, and M. Lewenstein, J. Phys. B {\bf 46}, 134010 (2013).

\bibitem{ngoldman2} N. Goldman, G. Juzeli\-{u}nas, P. \"Ohberg, and I. B. Spielman, arXiv:1308.6533.

\bibitem{huhui3comp} J. Chen, H. Hu, and Gao Xianlong, arXiv:1405.0709.

\bibitem{bcsbecreview} W. Ketterle and M. W. Zwierlein, {\it Making, probing and understanding ultracold Fermi gases}, Ultracold Fermi Gases, Proceedings of the International School of Physics ``Enrico Fermi'', Course CLXIV, Varenna, 20 - 30 June 2006, edited by M. Inguscio, W. Ketterle, and C. Salomon (IOS Press, Amsterdam) (2008).

\bibitem{chinchengreview} C. Chin, R. Grimm, P. Julienne, and E. Tiesinga, Rev. Mod. Phys. {\bf 82}, 1225 (2010).




\bibitem{fflo} P. Fulde and R. A. Ferrell, Phys. Rev. {\bf 135}, A550 (1964); A. I. Larkin and Y. N. Ovchinnikov, Sov. Phys. JETP {\bf 20}, 762 (1965).

\bibitem{nayak} C. Nayak, S. H. Simon, A. Stern, M. Freedman, and S. Das Sarma, Rev. Mod. Phys. {\bf 80}, 1083 (2008).

\bibitem{patrick} K. Michaeli, A. C. Potter, and P. A. Lee, Phys. Rev. Lett. {\bf 108}, 117003 (2012).

\bibitem{zhengzhenrashba} Z. Zheng, M. Gong, X. Zou, C. Zhang, and G.-C. Guo, Phys. Rev. A {\bf 87}, 031602(R) (2013).

\bibitem{Efimov} V. Efimov, {\it Yad. Fiz}. {\bf12}, 1080 (1970); Sov. J. Nucl. Phys. {\bf12}, 589 (1971).

\bibitem{halo1} M. V. Zhukov, B. V. Danilin, D. V. Fedorov, J. M. Bang, I. S. Thompson, and J. S. Vaagen, Phys. Rep. {\bf 231}, 151 (1993).

\bibitem{Braaten} E. Braaten and H.-W. Hammer, Phys. Rep. {\bf428}, 259 (2006).


\end{thebibliography}
\end{document}